\documentclass[12pt,preprint]{aastex}

\def\simless{\mathbin{\lower 3pt\hbox
   {$\rlap{\raise 5pt\hbox{$\char'074$}}\mathchar"7218$}}} 
\def\simgreat{\mathbin{\lower 3pt\hbox
   {$\rlap{\raise 5pt\hbox{$\char'076$}}\mathchar"7218$}}} 

\shorttitle{The MASIV Survey II}
\shortauthors{Lovell et al.}

\begin{document}

\title{The Micro-Arcsecond Scintillation-Induced 
Variability (MASIV) Survey II: The First Four Epochs} 

\author{J.~E.~J.~Lovell\altaffilmark{1,2,9}}
\email{Jim.Lovell@utas.edu.au}
\author{B.~J.~Rickett\altaffilmark{3}} 
\email{bjrickett@ucsd.edu}
\author{J-P.~Macquart\altaffilmark{4,10}} 
\email{jpm@astro.caltech.edu}
\author{D.~L.~Jauncey\altaffilmark{1}} 
\email{David.Jauncey@csiro.au}
\author{H.~E.~Bignall\altaffilmark{5,10}} 
\email{H.Bignall@curtin.edu.au}
\author{L.~Kedziora-Chudczer\altaffilmark{6}} 
\email{Lucyna.Kedziora-Chudczer@csiro.au} 
\author{R.~Ojha\altaffilmark{7}}
\email{rojha@usno.navy.mil}
\author{T.~Pursimo\altaffilmark{8}} 
\email{tpursimo@not.iac.es}
\author{M.~Dutka\altaffilmark{7}} 
\email{mdutka@umd.edu}
\author{C.~Senkbeil\altaffilmark{9}}
\email{cliffs@utas.edu.au}
\author{S.~Shabala\altaffilmark{9,11}} 
\email{stshabal@utas.edu.au}

\altaffiltext{1}{CSIRO Australia Telescope National Facility, PO Box 76,
  Epping NSW, Australia. Now at University of Tasmania}
\altaffiltext{2}{CSIRO Industrial Physics, PO Box 218 Lindfield NSW 2070, Australia}
\altaffiltext{3}{Department of Electrical and Computer Engineering, University 
of California, San Diego, La Jolla, CA 92093}
\altaffiltext{4}{NRAO Jansky Fellow, Department of Astronomy, California Institute 
of Technology, Pasadena CA 91125, U.S.A.}
\altaffiltext{5}{Joint Institute for VLBI in Europe, Postbus 2, 7990
AA, Dwingeloo, The Netherlands}
\altaffiltext{6}{School of Physics, University of Sydney, NSW 2006, Australia}
\altaffiltext{7}{US Naval Observatory, 3450 Massachusetts Avenue NW, Washington DC 20392, U.S.A.}
\altaffiltext{8}{Nordic Optical telescope, Sta Cruz de La Palma, E-38700 Tenerife, Spain}
\altaffiltext{9}{School of Mathematics and Physics, University of Tasmania, Tas 7001, Australia}
\altaffiltext{10}{Now at Curtin University of Technology, Bentley, WA 6845, Australia}
\altaffiltext{11}{Cavendish Astrophysics, J.J. Thomson Avenue, Cambridge CB3 0HE, United Kingdom}

\begin{abstract}
We report on the variability of 443 flat spectrum, compact radio sources
monitored using the VLA for 3 days in 4 epochs at $\sim 4$ month
intervals at 5 GHz as part of the Micro-Arcsecond Scintillation-Induced
Variability (MASIV) survey.  Over half of these sources exhibited 2-10\%
rms variations on timescales over 2 days.  We analyzed the variations by
two independent methods, and find that the rms variability amplitudes of
the sources correlate with the emission measure in the ionized
Interstellar Medium along their respective lines of sight.  We thus link
the variations with interstellar scintillation of components of these
sources, with some (unknown) fraction of the total flux density
contained within a compact region of angular diameter in the range
10-50$\mu$as. We also find that the variations decrease for high mean
flux density sources and, most importantly, for high redshift sources.
The decrease in variability is probably due either to an increase in the apparent
diameter of the source, or a decrease in the flux density of the compact
fraction beyond $z \sim 2$.  Here we present a statistical analysis of
these results, and a future paper will the discuss the cosmological
implications in detail.
\end{abstract}


\section{Introduction}

The discovery of centimeter-wavelength Intra-Day Variability (IDV) 
or ``flickering'' in  Active Galactic Nuclei (AGN) 
by \citet{hee84} initially raised concerns 
that some AGN possess brightness temperatures over six orders of magnitude 
above the $10^{12}\,$K inverse Compton limit for incoherent synchrotron emission 
\citep[e.g.][]{qui89}. However, considerable evidence has now accumulated 
to demonstrate that interstellar scintillation (ISS) in the turbulent, 
ionized interstellar medium (ISM) of our Galaxy is the principal mechanism 
responsible for the IDV observed in AGN, as was proposed by \citet{HR87}.

Two more recent observational techniques provide compelling evidence for the 
prevalence of ISS.  Time delays of 1--8\,min are observed in the arrival 
times of the flux density variations between telescopes on different 
continents for the three intra-hour variable sources B0405--385, 
B1257--326 and J1819$+$3845 \citep{jau2000,big2006,dtdb2002}.  The delay 
arises due to the finite time required for the stochastic fluctuations 
associated with the ISM to drift across the Earth.  A second observational 
signature of ISS relates to the modulation of IDV variability timescales 
with a period of exactly one year.  This arises because the Earth's 
orbital motion about the Sun contributes to the effective velocity with 
which the interstellar scattering material moves relative to an Earth-bound observer; 
the variations are slow as the Earth moves parallel to the material and 
fast as it moves anti-parallel to it. Annual cycles in IDV variability 
timescales are reported in at least seven sources 
\citep[e.g.][]{den2003,ric2001,jau2001,big2003,jau2003}, including several 
whose long variability timescales preclude detection of time delays in the 
scintillation pattern over intercontinental distances.  For many lines of 
sight through the ISM the slowest variations are expected in September if 
the motion of the turbulent material is comparable to the local standard 
of rest (LSR).

The recognition of ISS as the dominant cause of IDV has not entirely 
alleviated the brightness temperatures problems posed by these sources.  
A source must be small to scintillate; in the weak scintillation case most 
frequently observed at frequencies near 5 GHz \citep{wal98}, the source 
angular size must be comparable to or smaller than the angular size of the 
first Fresnel zone, $\theta_{\rm F}=\sqrt{c/2 \nu \pi L}$.  
Here $L$ is the distance to the scattering region, which we will refer 
to as the screen even though in some cases it may be better described as
a slab extending from the Earth out to distance $\sim L$.  
$\theta_{\rm F}$ is typically tens of microarcseconds 
for screen distances of tens to hundreds of parsecs, which
implies source components with angular sizes two to three orders of 
magnitude finer than the scales probed by VLBI. 
The long time-scale over which IDV has been observed in 
some sources suggests that such scintillating components can be relatively 
long-lived despite their small physical sizes.

This paper reports on the results of a Micro-Arcsecond 
Scintillation-Induced Variability (MASIV) survey for IDV in AGN. This 
year-long survey conducted observations of between 500--700 AGN over each 
of four epochs of three or four days duration in 2002 and 2003 at 4.9\,GHz 
with the VLA.
The aim of the survey was to provide a 
catalogue of at least 100 AGN which vary on timescales of hours to days to 
provide the basis of detailed studies of the IDV AGN population drawn from 
a well-defined sample.  A description of the observations in epochs 2, 3 
and 4 is presented in \S\ref{obs} as a supplement to descriptions of 
the first epoch observations and MASIV source selection in \citet{lov2003} 
(hereafter Paper 1). In \S\ref{VarClass} we describe how the time series of flux density for each source in each epoch was classified as variable or non-variable.  In \S\ref{DsClass} we describe how we have quantified the amplitude and timescale for the variations from an analysis of the structure function combined from all 4 epochs and apply corrections for noise and other sources of flux density error.  The basic hypothesis of the paper that the variations are predominantly due to interstellar scintillation is presented and examined in \S\ref{sec:ISS}, including the influence of parameters of the interstellar medium (\S\ref{VarB} emission measure and galactic latitude) and of the sources themselves (\S\ref{VarAlpha} mean flux density, spectral index).
We now have redshifts for more than half 
of the sources and we present the dependence of the variability on 
redshift in \S\ref{VarZ} - a result which shows that the MASIV survey 
provides a new cosmological probe.  In \S\ref{CompareVar}
we consider whether the variability is intermittent over the four epochs.  
Our data are listed in Table \ref{tab:BigTable} and
our conclusions are presented in \S\ref{Conc}.

\section{Observations and Data Reduction}
\label{obs}

The VLA observations took place over four periods during January, May and 
September 2002 and January 2003 (see Table \ref{tab:obs}). All epochs were 
72 hours in duration except September 2002, which included an additional 
24 hours.  The additional time in this epoch was added as an attempt to
detect the slower variation expected in September due to interstellar
scintillation caused by material moving at a velocity comparable to the LSR.

All observations took place during array reconfiguration and in each case 
the array was being moved to a more compact configuration (except for epoch \#1).  In each epoch 
the VLA was divided into five independent subarrays.  In the first epoch 
each subarray observed a subset of the 710 source sample.  Subarrays 1 
through 4 observed the ``core'' 578 sources of our compact, flat spectrum sample \citep{lov2003}, namely the weak 
($105<S_{\rm 8.6\,GHz} < 130$\,mJy) and strong ($S_{\rm 8.6\,GHz} 
>600$\,mJy) sub-samples, while the fifth subarray observed a sample of 
intermediate flux density sources in two regions of the sky.  For the next 
three epochs all sources previously observed in subarrays 1 through 4 were 
reobserved, thus providing full coverage for the core sample. However, 
subarray 5 was re-dedicated to observing a smaller number of sources 
comprising all objects found to be variable in subarray 5 during the first 
epoch, as well as the most rapid variables found during the first epoch 
which required faster sampling in order to determine time scales. In this 
paper we compare results for the core sample from subarrays 1 through 4 
only.

Shadowing was a serious concern at low elevations in the more compact VLA 
configurations. The impact of shadowing was minimized by 
assigning 5 or 6 antennas to each subarray in such a way that at any given time a 
source could be observed with at least 3 unshadowed antennas.  Any data from baselines containing shadowed antennas were flagged. Antennas that had 
either not moved during reconfiguration or were 
moved but had pointing solutions already applied were assigned 
to subarrays 1 to 4 where possible.

The core sample was divided into four roughly equal parts by Declination and a subarray assigned to each. The Declination ranges for the four subarrays were $0^\circ \leq \delta < 14^\circ$, $14^\circ \leq \delta < 34.08^\circ$, $34.08^\circ \leq \delta <  49^\circ$, $\delta \geq 49^\circ$. The Declination boundary between the second and third subarray was set to the latitude of the VLA to avoid long slews in azimuth when changing between sources transiting north and south of the zenith.

Each subarray was scheduled so that every source was observed for one
minute every $\sim$2 h while it was above an elevation of
$15^\circ$. We used the standard VLA frequency configuration for continuum 4.9~GHz observations (dual polarisation and two 50~MHz bandwidth IFs per polarisation) and a 3.3~s integration time.  For flux density calibration, each subarray
observed B1328+307 (3C286) and J2355+4950 every $\sim$~2
hours. B1328+307 is the primary flux density calibrator for the VLA and
J2355+4950 is a GPS source, not likely to vary over short timescales,
and is monitored regularly at the VLA as part of a calibrator
monitoring program.

Following the observations we calibrated the data in AIPS 
using the standard technique for continuum data. The task 
FILLM was used to load the data where corrections were 
made for known antenna gain variations as a function of 
elevation and for atmospheric opacity. Upon inspection 
of the data it was clear that there were residual 
time-dependent amplitude calibration errors.
We ascribe this to the fact that in each epoch, some VLA 
antennas had recently been moved and their
pointing calibration observations were not complete. The residual
pointing errors may depend on azimuth and
elevation so we chose several bright, non-variable sources in each
subarray at a range of RA as gain calibrators for
surrounding sources. Precautions were taken to ensure that the
calibrators themselves were not variable: if a given calibrator caused
the majority of sources against which it was applied to vary, then
another calibrator was chosen. These sources, typically 
recommended VLA calibrators, were drawn from our source sample.
The numbers of sources used as secondary calibrators in the four 
epochs were 42, 33, 20 and 36. 
On the epochs that a particular source was used 
as a calibrator, it is by definition non-variable and was excluded 
from the structure function analysis discussed below. 

Following calibration, the data for each source were 
inspected and occasional outlying samples were flagged. 
The data were then incoherently averaged on a one-minute 
timescale over all baselines. For all sources, the formal 
errors obtained were less than those estimated due to 
the residual constant and fractional errors discussed 
later in this paper. Incoherent averaging was chosen 
because, on the assumption that all sources are 
unresolved, the phase should be zero and any residual 
phase errors in the data would artificially reduce the 
average flux density. In low signal-to-noise regimes 
an incoherent average can induce an upward bias, as 
visibility amplitudes follow a Rice distribution, 
which has a mean that is systematically higher 
than the true mean \citep{tms2001}. In the case 
of our observations though, the signal-to-noise 
ratio of our observations is sufficiently high that 
this effect is negligible

As the VLA array configurations became more compact, our data became more 
sensitive to extended structure in the source and in nearby objects. The 
changing response of the VLA with time due to this structure can appear as 
variability in a light curve. Fortunately, as our observations were 
scheduled in sidereal time and were repeated at the same times every day, 
false variability due to structure appears as a repeating pattern with a 
period of one day. In general this was easy to recognise but in some cases 
imaging was carried out to verify contaminating structure. For the 
purposes of our analysis, such sources were removed from our sample. A 
total of 102 sources were removed due to structure or confusion,
and one was removed due to an error in our initial sample selection process.
Following these removals, we are left with a 
sample of 475 point sources common to all four epochs.

\section{Classification of Variables} 
\label{VarClass}

We used two separate approaches to analyze the flux density variations.  
This section describes how we classified the sources based on the 
apparent modulation index of their intensity.  The following section
describes a structure function analysis.  
The virtues of these two techniques are complementary: the first 
is based on a conservative but robust criterion of source variability, while 
the second uses a simple statistical estimator which
allowed us to quantify the amplitude and time scale of the variations.

In order to ascertain whether an individual source is variable it is 
necessary to understand the sources of error inherent to the measurements.  
There are uncertainties in the individual measurements
due to calibration errors causing a fractional error, $p$, 
and additive errors, $s$, due to thermal noise and confusion. 
Calibration errors include contributions from antenna pointing errors, 
system gain variation between the observations of flux calibrators and 
variable atmospheric absorption.  Since they are a small percentage 
of the mean source flux density, $\bar{S}$, they can be approximated as additive
and added in quadrature to the noise as given by
\begin{eqnarray}
\sigma_{{\rm err},s,p} = \sqrt{(s/\bar{S})^2 + p^2} \, .
\label{eq:error1}
\end{eqnarray}
Here $\sigma_{{\rm err},s,p}$ is the rms error in each flux density estimate 
normalized by the mean for each epoch. In our initial analysis (Paper I) 
we estimated $s=1.5$\,mJy and $p=0.01$. 

An initial inspection classified sources based on a
their modulation index, defined as the rms of 
the three days observations divided by the mean flux density, 
computed for each epoch.  A source 
was identified as variable if its modulation index exceeded 
twice the expected contribution from the measurement errors, $2\sigma_{{\rm err},s,p}$,
as in a $\chi^2$ test. 
However, direct inspection of the data revealed that many of the 
slower variables, that is sources with variability time-scales longer than 
three days, were not detected as variable using this criterion. 

We therefore introduced an alternative variability criterion, which 
classified a source as variable if the modulation index of its \it daily 
average \rm flux densities exceeded $2\sigma_{\rm err}$. This process yielded 
more detections, but visual inspection again revealed that many sources 
that clearly exhibited variability were undetected by either test.
In particular, examination revealed that our two criteria do not detect 
variability in those sources with low-level monotonic flux density changes 
during the 3-4 d duration of our observations.  This is because 
the $\chi^2$ statistic used in our two selection criteria is 
not an ordered statistic and is therefore sub-optimal in detecting such a 
low level trend.  

Our selection criteria were therefore augmented with a visual inspection of the remaining light curves. Inspection was performed by 
two of us independently and the results compared.  Each source was 
considered non-variable unless otherwise demonstrated, and any source 
where there was disagreement on its classification was reviewed; we 
adopted the conservative approach of classifying as non-variable any 
source is which no final agreement was reached on the classification. 
In the left panel of Figure \ref{fig:mraw} we show a scatter plot
of the raw modulation index $m_{\rm raw}$ of all sources against 
their mean flux density from epoch 1, using differing symbols for variables
and non-variables.  The predicted error in a single flux density measurement
from equation (\ref{eq:error1}) is shown by the dashed line,
which roughly separates the variable from non-variable classifications.
This emphasizes that our classification is determined largely by
$m_{\rm raw}$.  Table \ref{tab:BigTable} lists all the raw modulation indices and variability classifications of all the sources and Figures
\ref{fig:D1} and \ref{fig:lightcurves} show some sample 
light-curves from the survey.

In Table \ref{tab:SourceCounts} we list the number of sources classified
as variable in zero, one, two, three or all four observing epochs
and associated percentages.  In this table and in all of what follows we define
a set of 443 sources from the 475 sources that excludes those
used as secondary calibrators in two or more epochs. 
Note that 12\% of the sources were seen to vary during all four 
epochs.  With any analysis of a large number of observations, 
however, false positives, i.e. sources that are incorrectly classified as 
variable, are a significant concern. Since the visual classification was 
very close to the two sigma criterion, our variability 
classification is reliable with $\sim95$\% confidence.  Thus we give
in columns 3 and 4 the predicted fraction of sources misclassified 
based on the null hypothesis that all 443 sources are non-variable. 
Evidently these are negligibly small relative to the observed
fractions of variables in two or more epochs.

The large numbers of sources classified as variable on multiple epochs 
firmly establishes that our classification process is 
reliable.  If all 443 sources were non-variable, then the number 
expected to be classified variable on 2, 3 or 4 epochs is a 
mere 6.7 or 1.4\%, compared with the observed 192 (43\%). 
Even if our classification was 90\% 
reliable, then the number of false positives with 2 or more detections 
remains less than 5\%. 

The fraction of sources that can be reliably classified as variable 
can be deduced using Table \ref{tab:SourceCounts}.  Denoting $N$ as the 
actual number of non-variable sources, $T$ the actual number 
of one-time variables and $F$ the number of non-variables misclassified as one-time variables, 
then the apparent number of non-variables $161 = N-F$
and from Table \ref{tab:SourceCounts}, $F/N = 0.172$.
This yields $N = 194, F = 33$ and hence $T = 90-F =56$, and the corrected total 
number of variables is 192+56=248. Therefore the fraction of sources that exhibited variability 
in one or more epochs in our survey is 56\%. 
This value is significantly higher than 
the 15-20\% found in previous IDV surveys \citep{qwk1992,ked2001a}.
In comparing with other surveys one must be careful to consider
the selection criteria applied.  As described in \S\ref{obs} we
started with 710 sources selected from spectral index and mean
flux density criteria which was reduced to 443, when we excluded
those used as calibrators in more than one epoch and 
those exhibiting resolution effects, raising the percentage
of variable sources.   In addition the large number of one-, two- 
and three-time variable sources raises the percentage.  We
believe that this is due to intermittency in the IDV phenomenon,
which we describe by a simple model in \S\ref{CompareVar}.

For the purposes of subsequent analysis we conservatively define as ``non-variable''  those 161 sources that showed no variability in any of the four epochs, and we define as ``variable'' those 192 sources that showed variability on two or more of the four epochs. With these definitions we have two large 
and reliable samples each of approximately 200 sources, where the 
non-variables act as a control sample for the variables. Each was drawn 
from the same selection criteria and cover the same overall area of sky.

\section{Structure Function of the Variations} \label{DsClass}

Here we discuss how we quantify the flux density
variation for each source using the structure function (SF) of 
each time series, 
\begin{eqnarray}
D(\tau) =  \frac{1}{N_{\tau}} \Sigma_{j,k} (S_j - S_k)^2 
\label{eq:strfn}
\end{eqnarray}
where $S_j$ is a flux density measurement normalized by 
the mean flux density of the source over all four epochs and
$N_\tau$ is the number of pairs of flux densities
with a time lag $\tau$ binned in 2-hour increments.
The SF is a statistically reliable estimator 
which can be modeled even for short data spans. 
It is defined independent of any variability classification.
Examples are shown in the lower panels of Figure \ref{fig:D1}
and \ref{fig:lightcurves}. 
The SF is preferable to the autocorrelation function for short data spans,
which can be badly biased by a poor estimation of the mean.

For an idealized observation of \it stationary \rm stochastic variations
spanning a time much longer than their characteristic time $\tau_{\rm char}$, 
$D(\tau)$ rises with time lag and tends to saturation at twice the true variance.  
However for our observations $\tau_{\rm char}$ is typically more than two days and 
saturation is rarely seen.  We thus chose $D(\tau=2\,{\rm d})$
as a standard characterization of the intra-day variations 
because $\tau = 2\,$d is the maximum lag out to which our structure 
functions contain reliable information for the 72~hour observing sequences.   
Note that we estimate $D(\tau=2\,{\rm d})$ by combining data from all four epochs, 
which makes it a more stable statistic under the hypothesis
of stationary stochastic variations. Hence we do not use this statistic to 
determine whether a source's variability differs over the four epochs.

\subsection{Measurement errors}\label{measerr}

Additive flux density errors that are independent 
of the ISS (or intrinsic) variations contribute an additive term
to the SF.  If the errors are independent from one
sample to the next (i.e.\,``white''), the error contribution to SF 
would have a mean value $D_{\rm noise}=2 \sigma_{\rm err}^2$ independent of time lag.
While random pointing errors will be white, there may be non-white errors
caused by systematic pointing errors due to recent antenna relocations and 
residuals of gain and atmospheric variation, which may contribute a 
term that is a function of time lag.  Nevertheless we now proceed by assuming that all 
the errors are white and thus the errors add a constant ($D_{\rm noise}$)
(plus random variations about the constant due to estimation error) and postpone 
discussion of possible non-white measurement errors.

\begin{figure}
\includegraphics[width=15cm]{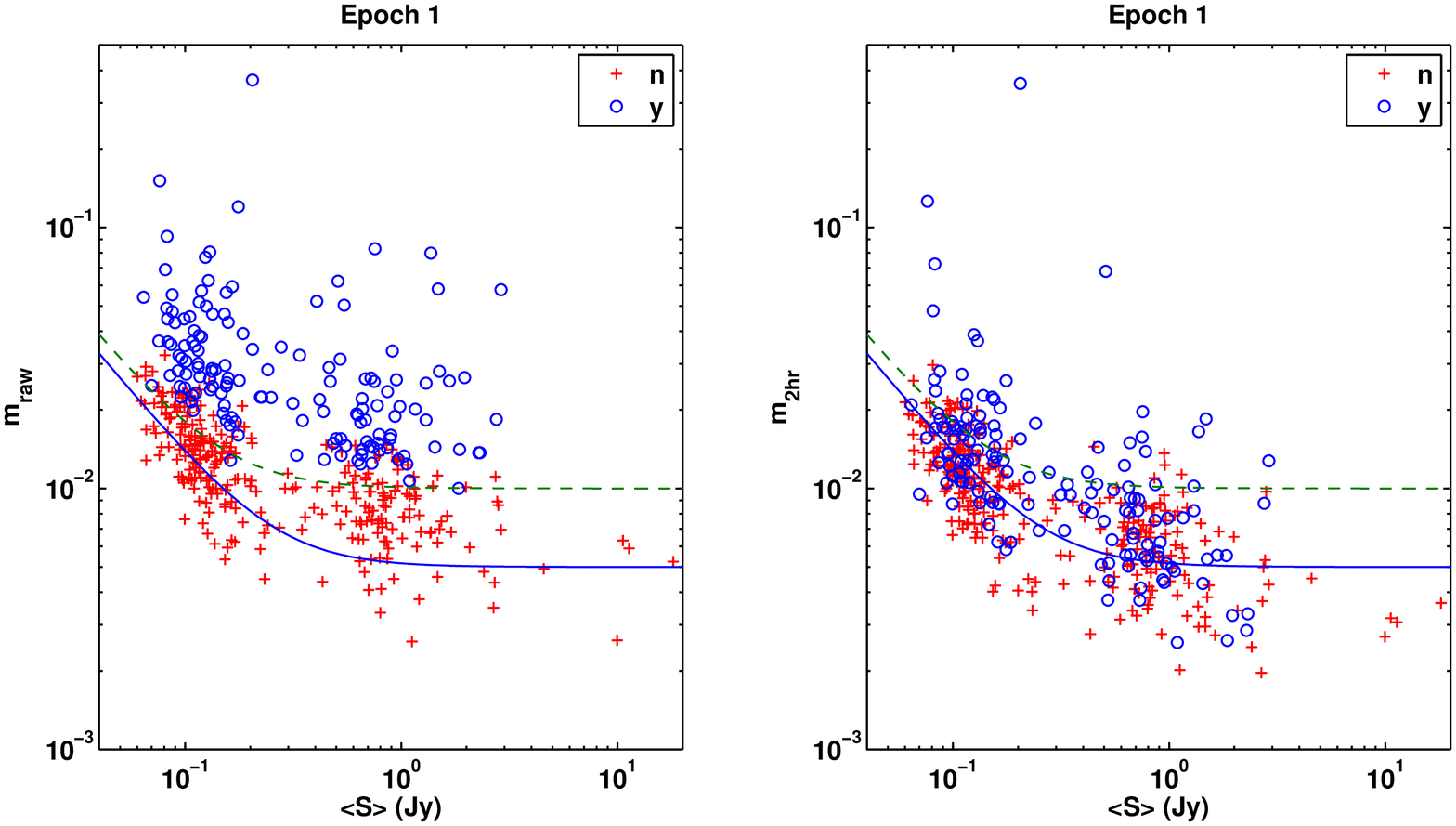}  
\caption{\it Left: \rm Raw modulation index 
for epoch 1 plotted against mean source flux density.
Sources classified as variable plotted as circles
and non-variable as pluses.
\it Right: \rm Similar plot for $m_{2{\rm h}} $ as described and defined in \S\ref{measerr}.
In both plots the dashed line is equation(\ref{eq:error1}) 
with $s=0.0015$ Jy and $p=0.01$
and the solid line has $s=0.0013$ Jy and $p=0.005$.
Similar plots are obtained for the other three epochs.}
\label{fig:mraw}
\end{figure}

\begin{figure}
\begin{tabular}{c}
\includegraphics[width=7cm]{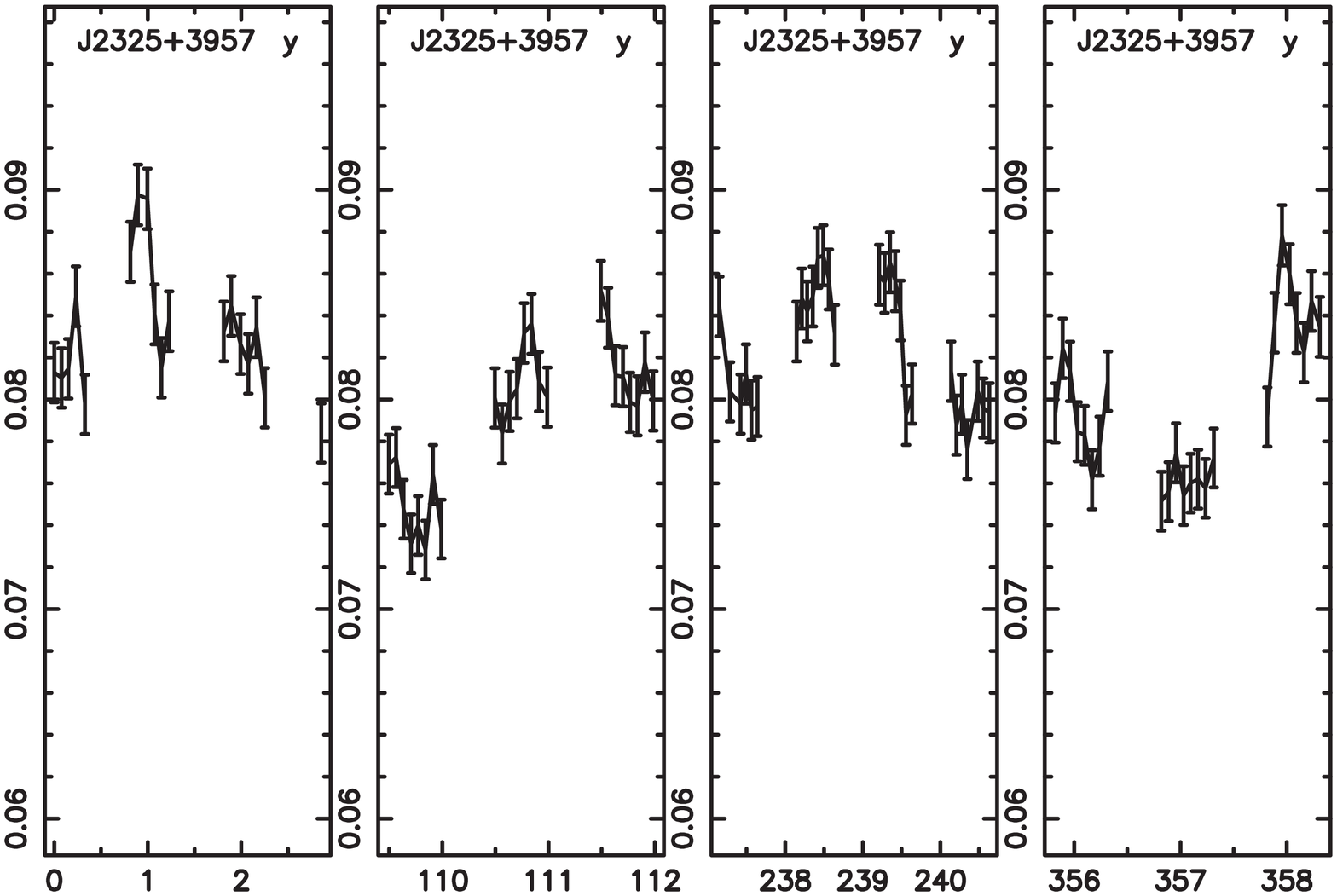} \\
\includegraphics[width=7cm]{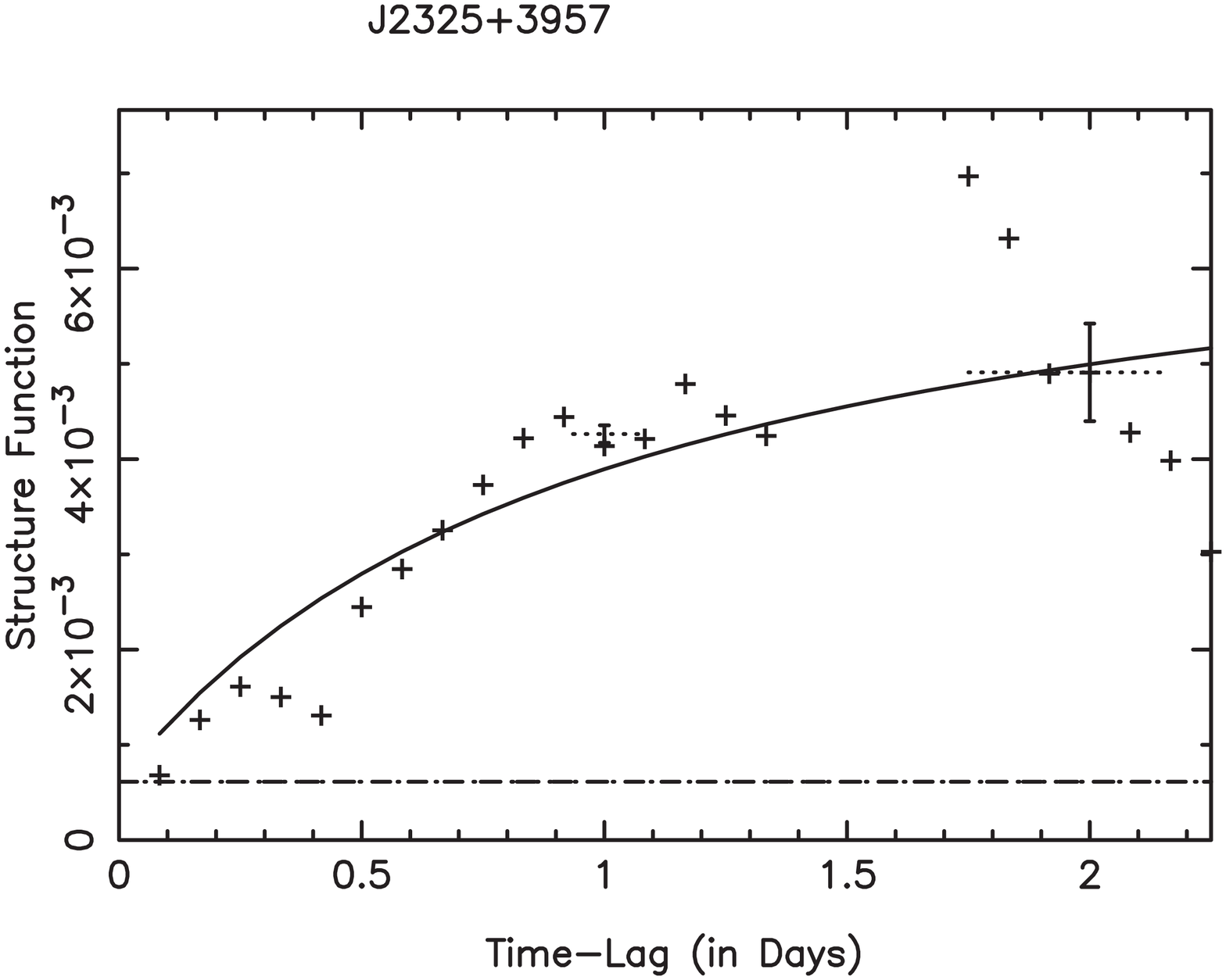} 
\end{tabular}
\caption{\it Upper: \rm Time series of flux density (Jy) for J2325+3957 versus
day number from 2002 January 1; error bars are as in Paper 1.   \it Lower: \rm
Structure Function $D(\tau)$ of flux density (normalized by its mean) 
averaged over all 4 epochs.  The dashed line is 
the estimated noise level $D_{\rm noise}$ and the solid line is a simple model fit 
(see \S \ref{VarTimeScale}).  The vertical bar centered near lag
of 2~d is an estimate of $D(\mbox{2\,day})$ and its standard error.
}
\label{fig:D1}
\end{figure}

\begin{figure}
\begin{tabular}{cc}
\includegraphics[height=10cm, angle=0]{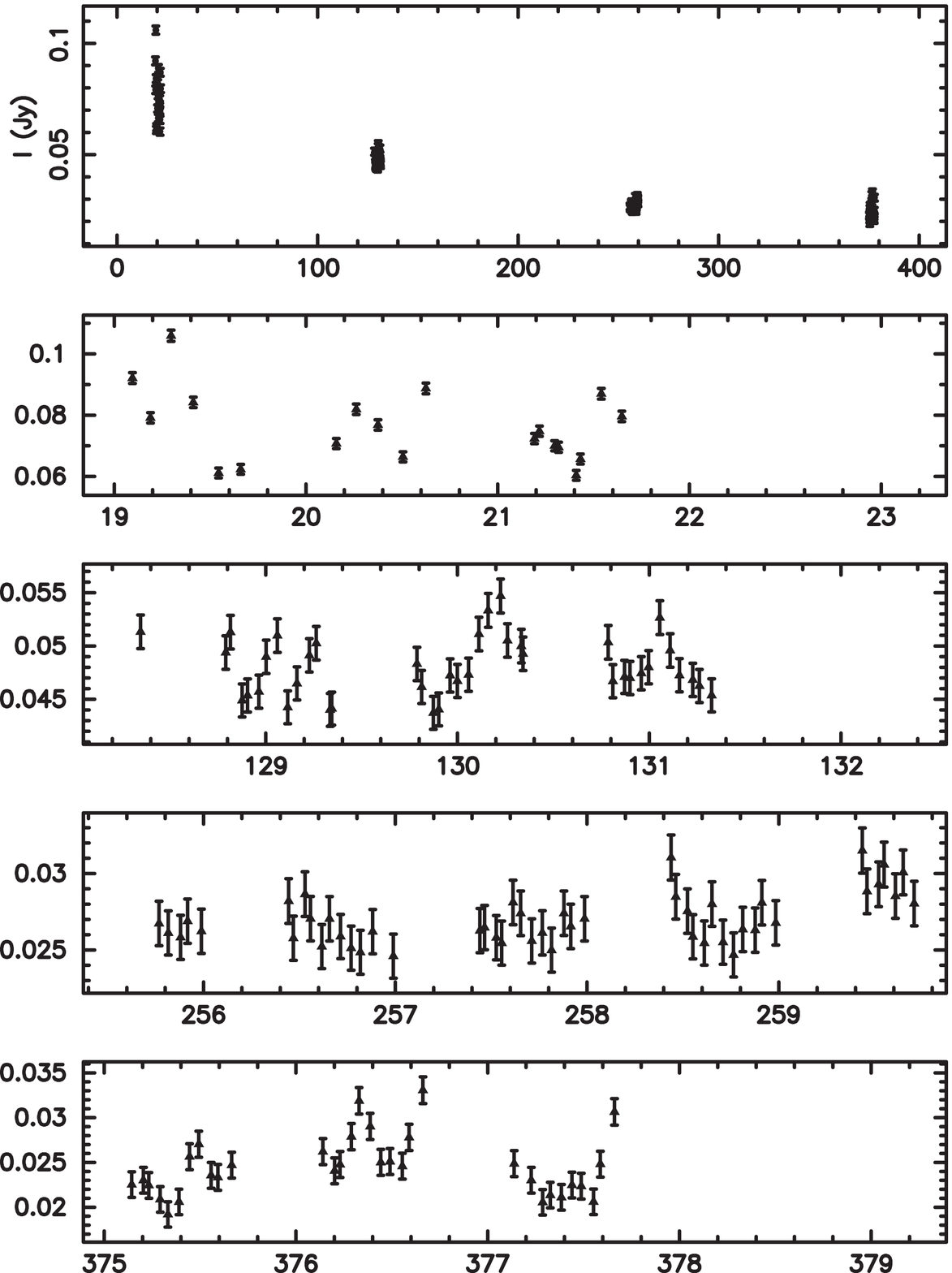} &   
\includegraphics[height=10cm, angle=0]{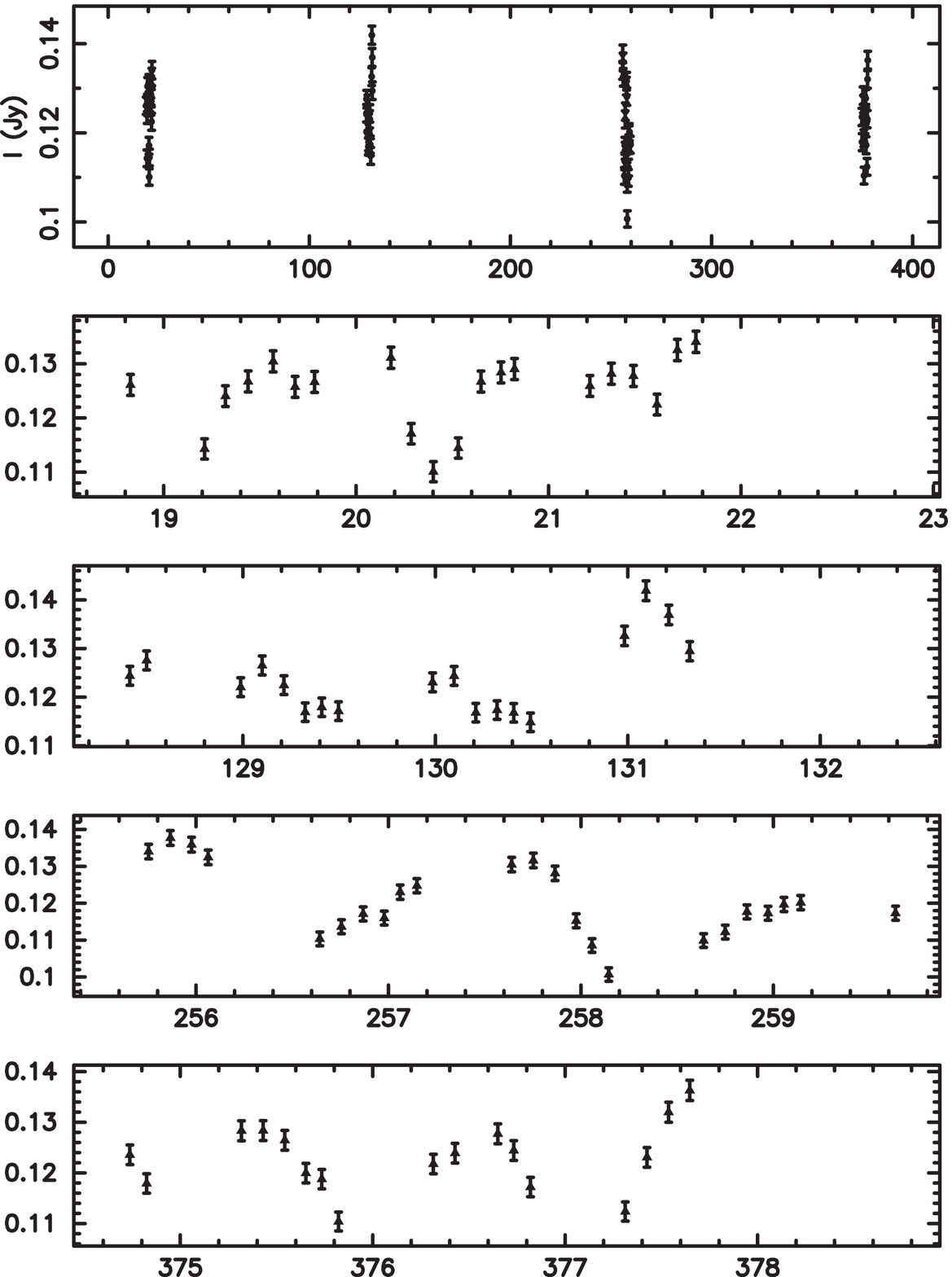} \\
\includegraphics[width=7cm]{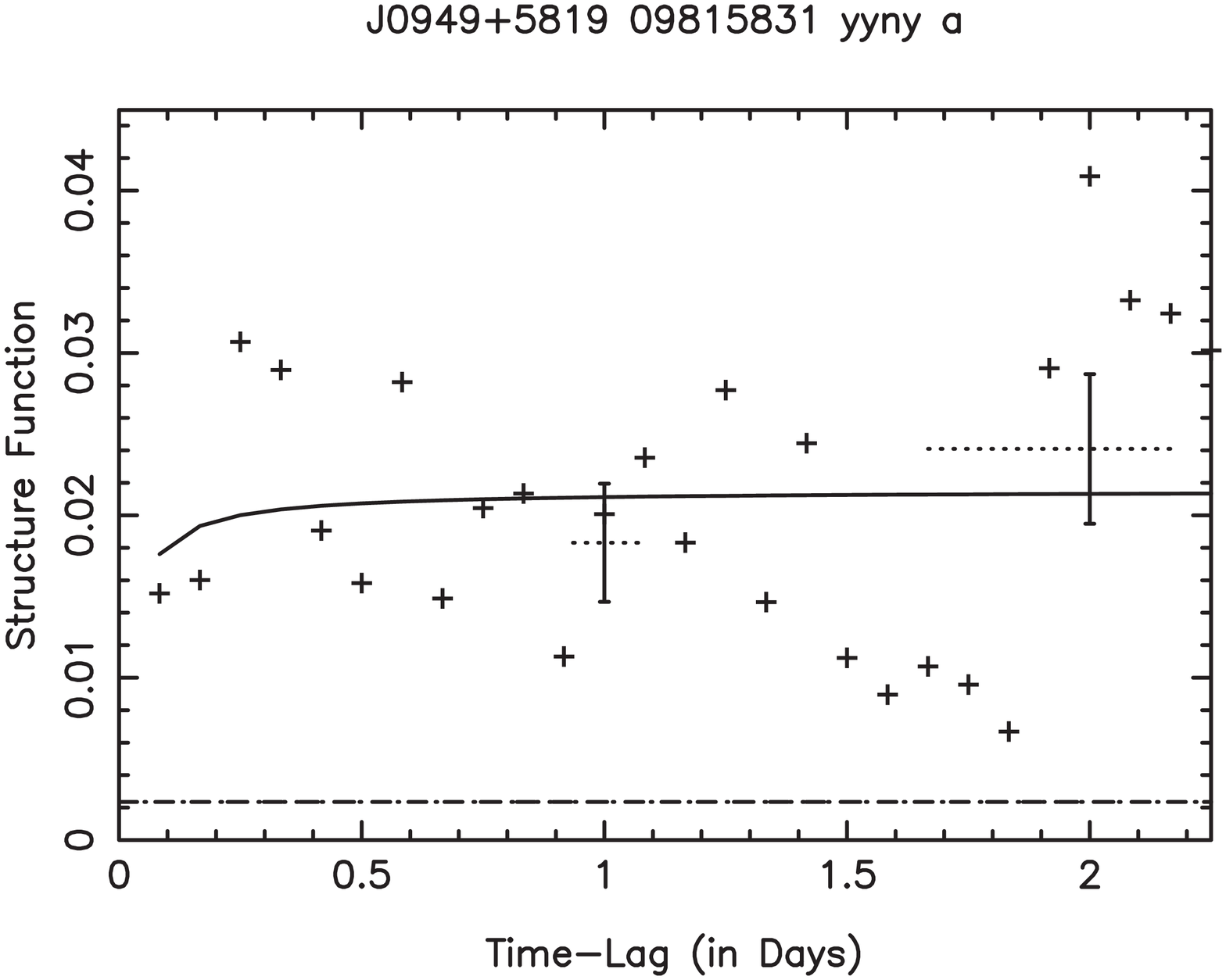} &
\includegraphics[width=7cm]{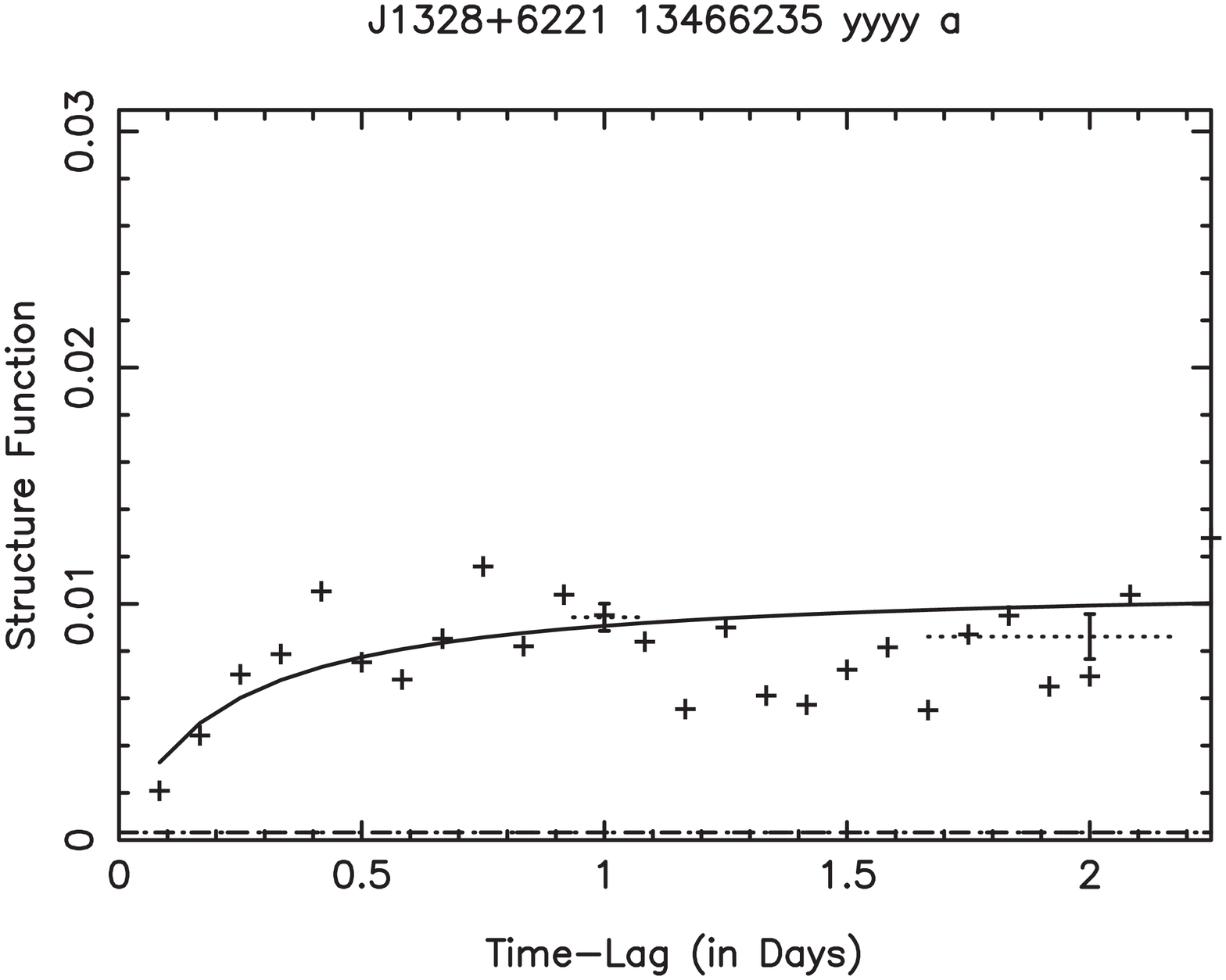} \\
\end{tabular}
\caption{Examples of MASIV variability. J0949+5819 (left) and
J1328+6221 (right). The top panel in each case shows all the
flux densities in Jy against day number from 2002 January
1. The middle four panels show light-curves for each of the first
four epochs. The horizontal scale is the same (four days) in each
case. Lower panels show structure function of flux density 
(normalized by its mean) averaged over all 4 epochs.  
Dashed line is the estimated noise level $D_{\rm noise}$
and the solid line is a simple model fit (see text).
J0949+5819 shows evidence of episodic scintillation, possibly related to intrinsic source changes as there appears to be a corresponsence between mean flux density over an epoch and amplitude of scintillation. J1328+6221 shows strong, consistent scintillation over all four observation epochs.} 
\label{fig:lightcurves}
\end{figure}

Our goal is to subtract $D_{\rm noise}$ from the raw SF 
$D(\tau)$ in order to estimate the SF of any true variations 
in a source's flux density.  But first we must consider how best to estimate
$D_{\rm noise}$. One estimate is $2\sigma_{\rm err}^2$ from equation 
(\ref{eq:error1}), which depends on the constants $s$ and $p$.  
Another estimate of $D_{\rm noise}$ comes from the SF 
itself -- evaluated at its shortest time lag (2h).  This would be an 
unbiased estimate for $D_{\rm noise}$ if all real flux density variation
had time scales much longer than 2 h.   Hence we examined single-epoch estimates
of $D(t=2{\rm h})$, which we plot as an equivalent modulation index 
$m_{2{\rm h}} = \sqrt{0.5 D(t=2{\rm h})}$ in the right panel of
Figure \ref{fig:mraw}.   The sources classified as variable include a 
substantial number with large values of $m_{2{\rm h}}$, which are
due to rapid flux density variations stronger than expected for noise.
Note in particular the highest
point which is quasar J$1819+3845$ which shows large amplitude
rapid ISS.  Its timescale is typically shorter than our 2~h 
sampling and so its SF has already saturated at 2~h
and its variations are ``white'' in our sampling.
Similarly the other sources with elevated $m_{2{\rm h}}$ are
probably due to ISS with short time scales.

Now consider the non-variables, which are plotted as pluses and
provide a set of sources with low or zero variation in epoch 1,
which are useful in studying the noise processes.
In the right panel of Figure \ref{fig:mraw} the mean of the plus symbols 
lies significantly below the dashed line
(equation (\ref{eq:error1}) with $s=0.0015$~Jy
and $p=0.01$) particularly for the higher flux density sources.  
The solid line with $s=0.0013$~Jy and $p=0.005$ provides a better 
model for the noise as discussed in \S\ref{sec:SF}.  
In the absence of any real variations
the estimates $m_{2{\rm h}}$ should be scattered equally above and below
their mean value. 
Comparing the right and left hand panels
we see that $m_{\rm raw}$ is typically higher than  $m_{2{\rm h}}$
even for the non-variables, which suggests that $m_{\rm raw}$
is increased due to low level variations with a timescale longer than
2~h.

The discussion above reduces the choice for estimating  
$D_{\rm noise}$ to either
$D(2{\rm~h})$ or the value corresponding to the solid line 
$D_{{\rm err},s,p} = 2 \sigma_{\rm err}^2$ with $p=0.005$.  Whereas it is 
appealing to use the observable $D(2{\rm h})$,  
any rapid real flux density variations  
can contribute to $D(2{\rm~h})$ which thus overestimates
$D_{\rm noise}$.  Thus we adopted $D_{{\rm err},s,p}$ 
as an estimator for $D_{\rm noise}$,
which states that the measurement errors are well described by 
equation(\ref{eq:error1}) but with calibration errors contributing 
at a 0.5\% rather than a 1\% level.  However we discuss a slight revision
of this in the next section.

\subsection{Structure Function Correction and Fitting}
\label{sec:SF}

For each source at each epoch we computed the raw SF as defined 
in equation (\ref{eq:strfn}).   As the time lag increases the number of
available lagged products ($N_\tau$) drops and so increases the error in the
SF.  A threshold for plotting an estimate for $D(\tau)$ was set  
at $N_\tau > 20$\% of the total number of data samples in that epoch.  
As the example in Figure \ref{fig:D1} shows, this gives estimates clustered 
at lags 2-8 h and near 24 h and near 48 h.
In order to characterize the variability amplitudes
we initially estimate the SF at a lag of 2\,d from the 
mean SF and its estimated error, calculated using the 
values at lags in the range $\tau = 48 \pm 2\,$hours.
The overplotted model is described in \S\ref{VarTimeScale}.

\begin{figure}[hbt]
\includegraphics[width=10cm]{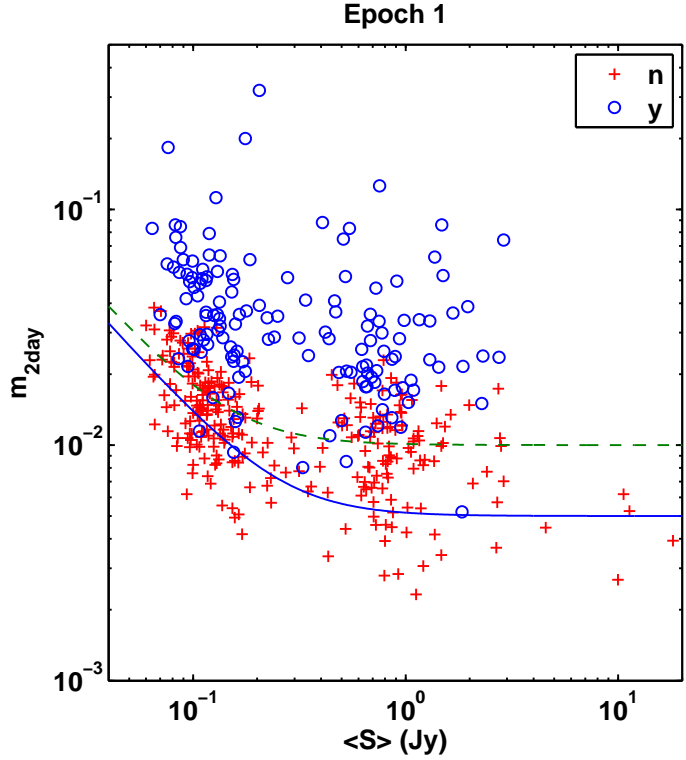} 
\caption{$m_{2\,{\rm d}} = \sqrt{D(2\,{\rm d})}$ 
(without any noise subtraction) plotted against mean 
source flux density.  Circles are sources classified as variables and 
pluses as non-variables.
The lines are equation(\ref{eq:error1}) with $s=0.0015$ Jy and $p=0.01$ for the dashed line and $s=0.0013$ Jy and $p=0.005$ for the solid line.
Similar plots are obtained for the other three epochs.
}
\label{fig:m2d1}
\end{figure}

First consider the equivalent 2-day modulation index
$m_{2\,{\rm d}} = \sqrt{0.5 D(2\,{\rm d})}$ (without noise correction)
plotted against the mean source flux density $\bar{S}$ for epoch 1
in Figure \ref{fig:m2d1}.  It should be compared with the 
left and right panels of Figure \ref{fig:mraw} for the non-variable
sources.  The comparison makes it clear that 
$m_{2\,{\rm d}} > m_{2\,{\rm h}}$ for almost all of the
non-variable sources.  Indeed there is a close correspondence between
$m_{2\,{\rm d}}$ and $m_{\rm raw}$.  Thus it is clear that there are
low level flux density variations on a timescale longer than 2~h
in the sources classified as non-variable.
In order to investigate what causes these we examined the difference 
$\Delta D = D(2\,{\rm d})-D(2\,{\rm h})$ for the \it non-variable \rm 
sources from each epoch.

We examined how $\Delta D$ depends on mean flux density, Galactic latitude
and H$\alpha$ emission -- quantities which we find 
in \S\ref{sec:ISS} influence $D(2\,{\rm d})$ for the 
\it variable \rm classifications.
The results showed that the mean of $\Delta D$ in each epoch
was significantly higher for the sources weaker than
0.4 Jy than for those stronger than 0.4 Jy.   However, it
showed no significant dependence on Galactic latitude or H$\alpha$
emission.  Hence the process
responsible for the low-level variability in the sources classified as
non-variable is not ISS and is unlikely to have an astrophysical origin.  

We consider the most likely cause to be confusion, which
can be due either to extended source structure which is partially resolved 
on the baselines of each sub-array or to low level confusing sources in the 
primary beam.  We had already eliminated obvious cases of confusion by 
removing sources whose light curves showed clear daily patterns of 
variability which repeated in each day of a 3 day sequence.  Since our SF estimation is from
the light curves normalized by $\bar{S}$ an increase in $\Delta D$
at low $\bar{S}$ is consistent with the effect of low level confusion
characterized by a certain rms in Jy.  This is supported by the VLA 
documentation that gives 2.3~mJy as the brightest source expected
in a single antenna beam at 5 GHz.

We examined the shape of the 
mean $D(\tau)$ for high and low $\bar{S}$ sources in each epoch.
Evidence for the effects of confusion was found in the 
consistent minima in $D(\tau)$ near time lags $\tau =$ 1d and 2d.
However, there was also a component rising from the noise floor at 2h
and typically saturating between 12 and 24h.  All three components
were substantially larger for the weak group of sources
than for the strong sources.  When averaged over four epochs
the noise floor was, respectively, at
$\sim 3.0\times10^{-4}$ and $\sim 0.61\times10^{-4}$  
and the averages for $\Delta D$ were, respectively, 
$\sim 2.8\times10^{-4}$ and $\sim 1.1\times10^{-4}$
for these groups (mean flux density 0.13 Jy and 1.4 Jy). 
By equating these average SF amplitudes to $2\sigma_{err}^2$ using equation
(\ref{eq:error1}) we estimated $s=0.0013$ Jy and $p=0.0073$.  
The $s$ values for each epoch  ranged from
0.001-0.002 Jy and for $p$ 0.003-0.01, but no consistent patterns
were seen versus epoch.

We also considered the effect of long-term variations as characterized by
the variation between epochs of the mean flux density from each epoch.
These slower variations are mostly intrinsic and might contribute
to a trend within the 3 or 4 days of each epoch.  So we calculated
the rate of change of flux density from
the average of the magnitude of the differences in flux density
between neighbouring epochs, divided by the number of days between 
epochs and the resulting magnitude of $D(2{\rm d})$ due to such a trend.
Surprisingly, for all sources this was smaller than the noise-corrected $D(2{\rm d})$  
except for those which were negative as a result of noise subtraction.  
Further the highest of them was 0.0001, which is one quarter of
the threshold value. So we conclude that long-term variations did not make a significant 
contribution to the variations observed within the epochs.  We note,
however, that we can usefully estimate the structure function on 
time lags of 3 and 6 months from the MASIV survey data, which 
may be useful in studies of intrinsic variability.  
Thus we include the epoch averaged
flux density for each epoch in the data table \ref{tab:BigTable}.

The foregoing studies reveal that the apparently non-variable light curves
include not only white noise but also a low level contaminating process
whose rms values can be approximately characterized 
by equation (\ref{eq:error1}).   While we can
quantify the white noise process by this equation, the non-white contamination
is not well enough understood to be reliably characterized by an
SF which could be subtracted from the estimates for all sources at each epoch.
So in order to reduce the effect of these non-white
variations, we increased the estimated $D_{\rm noise}$ to
to be subtracted by using the values $s=0.0013$ Jy and $p=0.007$,
and we also set a threshold on $ D(2\,{\rm d}) = 4\times 10^{-4}$ 
below which the noise-corrected SF values may be contaminated 
and therefore should only be interpreted as 
upper limits to the true SF of flux density variations.

It is of interest that, since confusion effects will be precisely repeated
at 24~h intervals, the samples of the structure function at
24 and 48~hr will be unaffected by confusion.  The results we present in
subsequent sections are from fitting the SF over all time lags, since
the fit makes better use of the data.  However, we also analyzed the
single sample estimates of $D(24{\rm h})$ and $D(48{\rm h})$
and found very similar results, though with somewhat worse
statistical errors.  As an extra precaution we re-reviewed 
all the light curves and SF plots for 24~h periodic patterns
and found 34 sources that might be contaminated at a level near
the threshold.  For these sources we
used the lower of $D(2\,{\rm d})$ from the fit and
that estimated from $48\pm2$~h.

\subsection{Variability Timescales} 
\label{VarTimeScale}

Although we estimated the SF as described above for each source at each epoch,
the single-epoch $D(2\,{\rm d})$ is only based on about one independent
sample of a 2~d variation and so it has a large statistical error
-- such that its rms error is about equal to its mean
\citep{RCM2000}.
Thus we do not attempt to evaluate the variability amplitude
on a source by source basis for each of the epochs.
Rather we average the SF for each source
over all 4 epochs (after subtracting the $D_{\rm noise}$ as defined
in the previous section).
Hence the SF results are insensitive to any intermittency or
annual changes in ISS due to the effects of the Earth's velocity.

We then fitted the following simple model to the SFs:
\begin{eqnarray}
D(\tau) = 2 m^2  \frac{\tau}{\tau + \tau_{\rm char}},
\label{eq:dmodel}
\end{eqnarray}
where $2 m^2$ is the amplitude at which the function 
would saturate and $\tau_{\rm char}$ is the characteristic
timescale where the SF reaches half of its saturation.
The motivation for this form is described 
in Appendix \ref{DsISS}. It approximates the form 
expected from ISS caused by a turbulent interstellar 
medium uniformly distributed through a thick scattering 
region, as opposed to turbulence confined to a thin layer 
(see equation(\ref{eq:Dmodel})).  We estimated two parameters 
only: the timescale $\tau_{\rm char}$ and the value of 
$D(\tau=2\,\mbox{d})$ for each source.  It should also
be noted that a light curve that is dominated
by a linear trend in flux density gives rise to 
a parabolic SF, which is not well fitted by equation
(\ref{eq:dmodel}).  The value of $D(\tau=2\,\mbox{d})$
in such a case will be somewhat underestimated.

An example of fitted SF is shown 
in Figure \ref{fig:D1}. 
The points show $D(\tau)$ increasing
(noisily) with time lag $\tau$ but not reaching saturation.
Since the timescale is defined at half the saturation value
it is poorly constrained in this example: 
$\tau_{\rm char} = 1.0 \pm 0.5$~days.  
However, from the same fit $D(t=2\,\mbox{d}) = (44 \pm 3)\times 10^{-4}$, 
which is quite well constrained.  With observations 
limited to 3~days (4~days for epoch 3) it is not 
possible to estimate timescales longer than about 3~days.  
However, in those cases it was possible to recognize that 
the characteristic timescale is longer than 3~days from the
shape of the structure function.  
Two other examples are shown in 
Figure \ref{fig:lightcurves}, in 
which there is evidence for faster variations.  
For source J0949+5819 the variations are very strong in epoch 1
and much weaker in epoch 3.  From the epoch-average 
we find $\tau_{\rm char} = 0.02 \pm 0.05$~days and 
$D(t=2\,\mbox{d}) = (19 \pm 2)\times 10^{-3}$.
For J1328+6221 the variations are more consistent
over the epochs with 
$\tau = 0.2 \pm 0.2$~days and 
$D(t=2\,\mbox{d}) = (10 \pm 1)\times 10^{-3}$.  
Given our 2~hr sampling and the typically large fractional errors
in the timescale we have simply classified the timescale into fast 
$\tau_{\rm char} < 0.5$\,d, medium $0.5 <\tau_{\rm char} 
< 3$\,d and slow $3\,{\rm d} < \tau_{\rm char}$. 
We also looked for any correlation between the
timescale and $D(t=2\,\mbox{d})$ but found no consistent pattern.

\begin{figure}[!ht]
\begin{tabular}{c}
\includegraphics[width=9cm,angle=270]{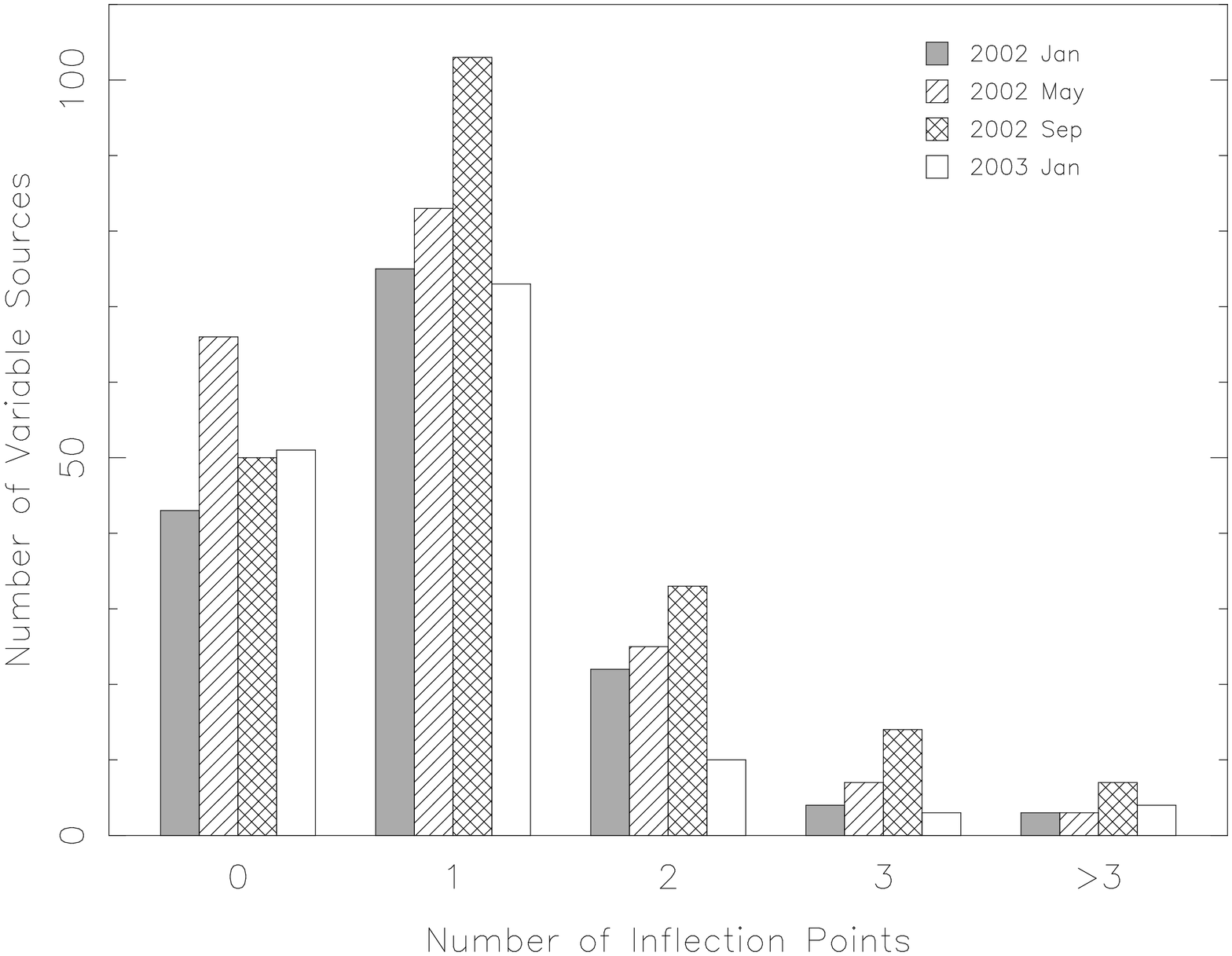} \\
\includegraphics[width=9cm,angle=270]{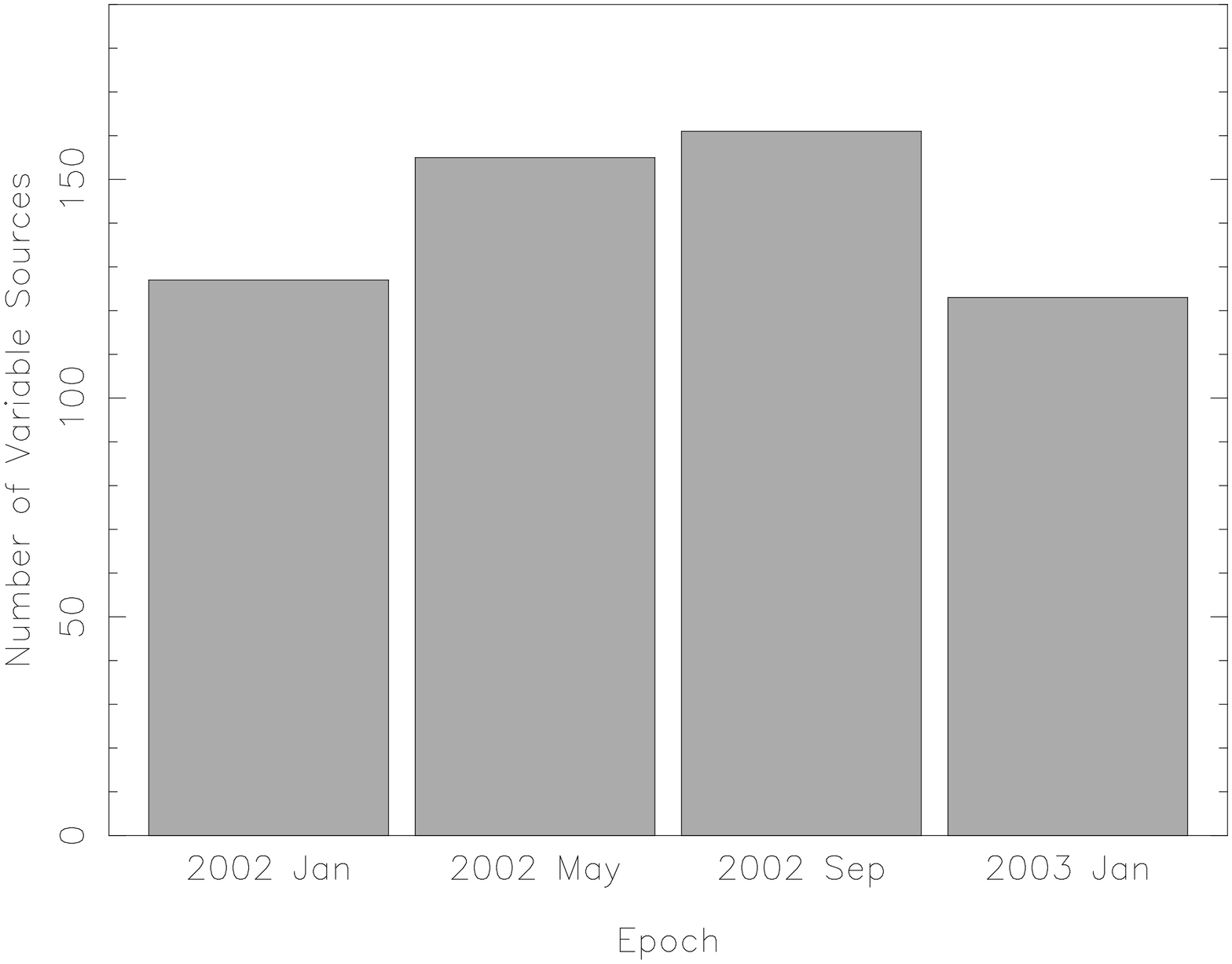} \\
\end{tabular}
\caption{\it Upper: \rm The number of sources classified as variable versus the
observed number of changes in the sign of the derivative of flux
density vs time (i.e. number of inflection points). Clearly a
majority of the sources vary on timescales of 3 days or longer.
\it Lower: \rm Numbers of sources classified as variable in each
epoch. Screens moving at the LSR would be
expected to result in fewer variables beeing seen in epoch 3; we see
no such deficit. }
\label{InflectionTimescale}
\end{figure}

During the visual examination of the light curves for
each source at each epoch, the timescales were estimated
by counting the number of inflection points
(i.e. change in sign of the derivative) for those epochs
classified as variable.  Since in the visual examination
there was an effective smoothing, inflection points due to noise-like
deviations were not counted.
The majority of sources were found to show none or at
most one inflection point indicating variability timescales that are
predominantly longer than 3 days. The observed distribution of
inflection points is shown in Figure \ref{InflectionTimescale}. 
Only a small number of sources (20\%) showed 2 or more inflection points. 
A comparison of the distribution of inflection points for the weak and
strong sources revealed no significant difference between the two
classes.  Overall, the distribution of
timescales was statistically the same for each epoch, remembering that
epoch 3 was four days rather than the three days of the other epochs.
An important conclusion from the timescale study is
our 3 or 4~d lightcurves commonly underestimate 
both the timescale and true modulation index for many
of the sources.

The annual cycle reported in a number of IDV sources is due to the changing
relative velocities of the Earth and the ISM responsible for the
scattering \citep{mac2003}. If the ISM velocities follow the Local
Standard of Rest, many sources would be expected to 
exhibit slower variations in the third quarter of the year 
(i.e. during the third epoch), and hence may more easily be missed because 
of the lengthened time-scales.  Figure \ref{InflectionTimescale} shows the numbers of
variables found at each of the four epochs. A Chi-squared contingency test shows
no evidence that the numbers differ from the mean in any epoch, even
though epoch 3 lasted for four rather than three days. The uniformity
of variable numbers in each epoch suggests a lack of evidence for a
third-quarter slow-down, and it follows that the majority of the
scattering material is not moving at the LSR. This is perhaps not
unexpected; both PKS~1257--326 and J1819+3845, the two sources for
which reliable screen velocities have been measured, have measured
screen velocities that differ significantly from the LSR 
\citep{big2006,den2003,lrr2007}.

In summary we used the visual analysis to classify each source at each epoch
as variable or not variable.  We computed SFs for
all sources and then examined the SFs of those classified as 
non-variable in order to quantify the measurement errors.
We were able to correct the SFs by subtracting a constant versus 
time lag due to errors that are independent over the 2~h sampling
and are characterized by equation \ref{eq:error1}. In addition
we found a low level contaminating process with a timescale of 1-2 d, 
which we suggest is due to low-level confusion with an rms of 1-2 mJy.  
The SF of the slower contamination could not be reliably estimated
and so sets a limit on the minimum detectable variation in flux density.
By fitting a simple curve to the epoch-averaged and noise-corrected SFs,
we estimated $D(t=2\,\mbox{d})$ for each source. 
The contamination is minimized by requiring
this quantity to be above a threshold of $4\times10^{-4}$
for a useable estimate of the timescale $\tau_{\rm char}$.  
Most sources were classified as slow variables.

\subsection{Comparison of SF with the visual variability 
classification}

\begin{figure}[hbt]
\includegraphics[width=13cm]{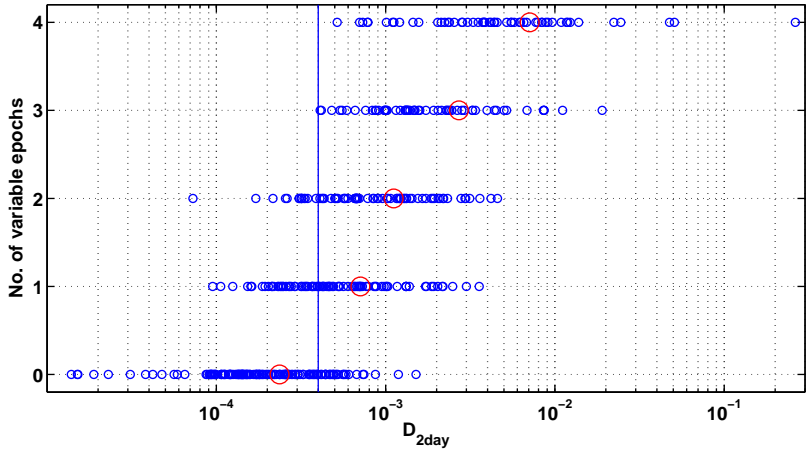} 
\caption{
The number of ``variable'' epochs plotted against
the value of $D(2\mbox{d})$ fitted to the cumulative
structure function for all four epochs. 
The large circles show the mean values for each group of sources
(see Table \ref{tab:compvar}).
}
\label{fig:nevar}
\end{figure}

We now compare the SF analysis with the 
variability classification from \S\ref{VarClass}.
Figure \ref{fig:nevar} shows the number of epochs in which a source
was classified as variable plotted against its value of $D(2\,\mbox{d})$
obtained as described from a fit to the structure function 
of the cumulative data from all 4 epochs. 
The large circle shows the mean values for each group of sources.
While it is clear that the sources with higher  $D(2\,\mbox{d})$
were classified as variable more frequently, there is a very wide
distribution in the rms level of the variation over 2~d 
among the sources. The vertical line marks the threshold 
$D(2\,\mbox{d}) = 4\times 10^{-4}$ above which we
have made a timescale estimate.  Values below this should be regarded as upper
bounds in view of the possibility of low level confusion.
 
Table \ref{tab:compvar} lists the source counts
sorted by the number of ``variable'' epochs for the 443 sources
(as always excluding those used as calibrators in more than one epoch.)  
It also shows the mean values of $D(2\,\mbox{d})$ and the numbers of sources
above and below the SF threshold. In total 
37\% of them are above the threshold versus 
45\% from the variability classification on 2 or more epochs.
In the latter classification process we do not characterize
the rms amplitude but attempt to quantify any intermittency
in the phenomenon.  In contrast the SF analysis quantifies the 
rms amplitude over 2~d averaged over all 4 epochs.

\section{Interpretation as Interstellar Scintillation}
\label{sec:ISS}

We now examine our basic hypothesis, stated in the introduction, that 
the variations in flux density detected in the MASIV survey are caused 
by interstellar scintillation (ISS).   The extremely small diameters of pulsars
revealed the ISS phenomenon almost as soon as pulsars were discovered
and they still provide the best information on the distribution of 
small scale structure in the ionized interstellar medium --
indeed they provide a calibration of the ISS phenomenon.  
The pulsar observations have been combined into a model for the distribution
of electron density in the interstellar medium by \cite{tc93} and 
revised by \cite{cl05}.  
  
Because the refractive index of an ionized medium varies with radio frequency,
there is a transition frequency ($f_{\rm w}$) above which the scintillation of
a point source, like a pulsar, is weak in the sense that its scintillation
(modulation) index ($m_{\rm pt}$) is less than one.  This frequency 
is on the order of 5~GHz but depends on the strength of the turbulent fluctuations
in electron density on the given line of sight \citep{wal98,cl05}.  Above $f_{\rm w}$ the ISS of a point source has a single 
timescale approximately given by $t_{\rm F}= r_{\rm F}/V$, 
where $V$ is the effective velocity of the Earth through
the ISS diffraction pattern and $r_{\rm F}=\sqrt{L\lambda/2\pi}$
and $L$ is the typical distance to the scattering region.

In our observations the angular diameters of the extra-galactic sources
are considerably larger than those of pulsars and so their ISS is heavily 
quenched. See \citet{ric2006} for a discussion of how the ``low-wavenumber 
approximation'' can be applied for quenched scintillation even below the
transition frequency.  If we approximate the scattering as taking place in a thin
region of the Galactic plane, we obtain simple expressions for the reduction
in scintillation index and lengthening of timescale  (e.g. \citet{R86}).
In Appendix \ref{DsISS} we apply the same simple ``screen'' model to the 
structure function analysis and obtain expressions for how the 
observable $D(2\,\mbox{d})$ might vary with angular size of each source 
and on distance to the scattering region and the level of turbulence on 
each line of sight.  Of course the level of ISS also depends on properties
of the source -- in particular the fraction of its flux density 
in its most compact component and on the effective diameter of that component.

\subsection{Galactic Dependence of ISS} \label{VarB}

We start by comparing our $D(2\mbox{d})$ results
with the emission measure (column density of the square of the electron density)
as estimated from observations of H$\alpha$ emission.
We find the intensity of H$\alpha$ 
emission (in Rayleighs) from the WHAM Northern sky survey on a 1 
degree grid \citep{wham} nearest to each source.  
We use the intensity summed over all velocities,
which \citet{wham} interpret as proportional to the ISM emission measure
on that line of sight, assuming the temperature of the emitting gas
does not vary by a large percentage.  We expect the level of ISS
to be related to the emission measure on that line of sight, as described
by \citet{sc98} and observed by \citet{ric2006}. 

\begin{figure}[hbt]
\includegraphics[width=13cm]{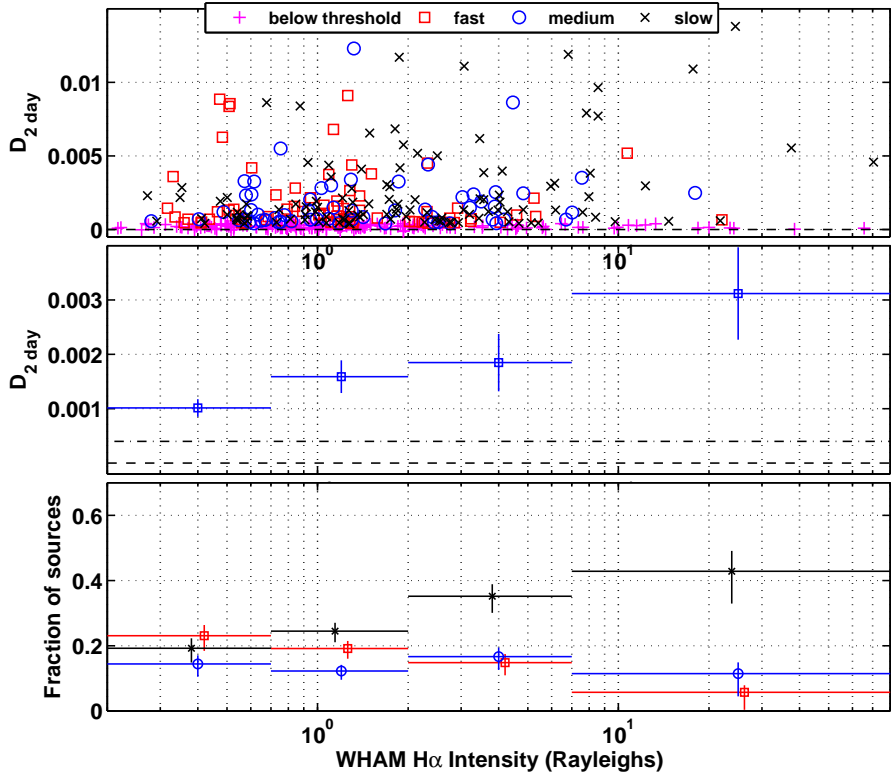} 
\caption{\it Upper panel: \rm Log-log Scatter plot of $D(2\mbox{d})$
against WHAM H$\alpha$ emission which is proportional
to the emission measure on a line of sight sampled on a 1 degree grid 
of the Northern sky \citep{wham}.  The intra-hour variable
source J1819+3845 is off scale at 0.25.
The different symbols represent the
three classifications of ISS timescale, as described in the text.
\it Center panel: \rm Mean value of $D(2\mbox{d})$ in
the indicated bins of H$\alpha$ emission including 
all sources except J1819+3845; vertical bar gives the standard error in the mean. 
\it Lower panel: \rm  Fraction of sources above the threshold in each timescale 
class in each bin showing that fast ISS is commonest
for the lower column density of electrons and slower ISS
dominates for higher column densities.  Error bars assume binomial 
distributions.  The same method is used in Figures \ref{fig:latplot},
\ref{fig:sid2d}, \ref{fig:saveplot} \& \ref{fig:zplot}.
}
\label{fig:Halpha}
\end{figure}

Figure \ref{fig:Halpha} plots $D(2\mbox{d})$ against the WHAM H$\alpha$ 
emission (in Rayleighs).  Though the scatter plot in the top 
panel shows little obvious trend, 
the bin averages in the middle panel show a clear upward
trend with emission measure, which establishes ISS as the 
dominant cause of the variability in the MASIV survey.  
We stress  the complete independence of 
the two data sets in this figure and that the bin averages are 
independent of any threshold set on $D(2\mbox{d})$.
We exclude the extreme IHV source J1819+3845
from this and subsequent bin average plots because its $D(2\mbox{d})$
value is 0.25 which is so much higher than the next highest at 0.015
that it distorts the mean and the variance within its bin. 

The bottom panel shows that
the fraction of slowly scintillating sources clearly increases
with  emission measure and vice-versa for the fast scintillators.
This finding that the longer timescales occur when seen through 
greater column density of electrons is consistent with 
enhanced ISS from strongly ionized regions of the ISM, which are typically
at low Galactic latitudes and at greater distances $L$.
An increase in $L$ increases the scale of the scintillation
pattern which slows the scintillation time - see Appendix \ref{DsISS}.

Since the strength and effective distance of the scattering layer
depends on Galactic latitude, we also expect a dependence of ISS on
Galactic latitude.  For our visual classification of sources 
(as variable or non-variable) we asked the
simple question ``are their latitude distributions the same?''.
A Chi-squared contingency test dividing the sources into two
samples, a low latitude sample, $|b| <$ 40 degrees, and high latitude
sample, $|b| >$ 40 degrees, shows that the two distributions differ at 
the 98\% confidence level. There are fractionally
more variables at low latitudes than there are at high latitudes, supporting
ISS as the origin of the intra-day variability. 

\begin{figure}[hbt]
\includegraphics[width=13cm]{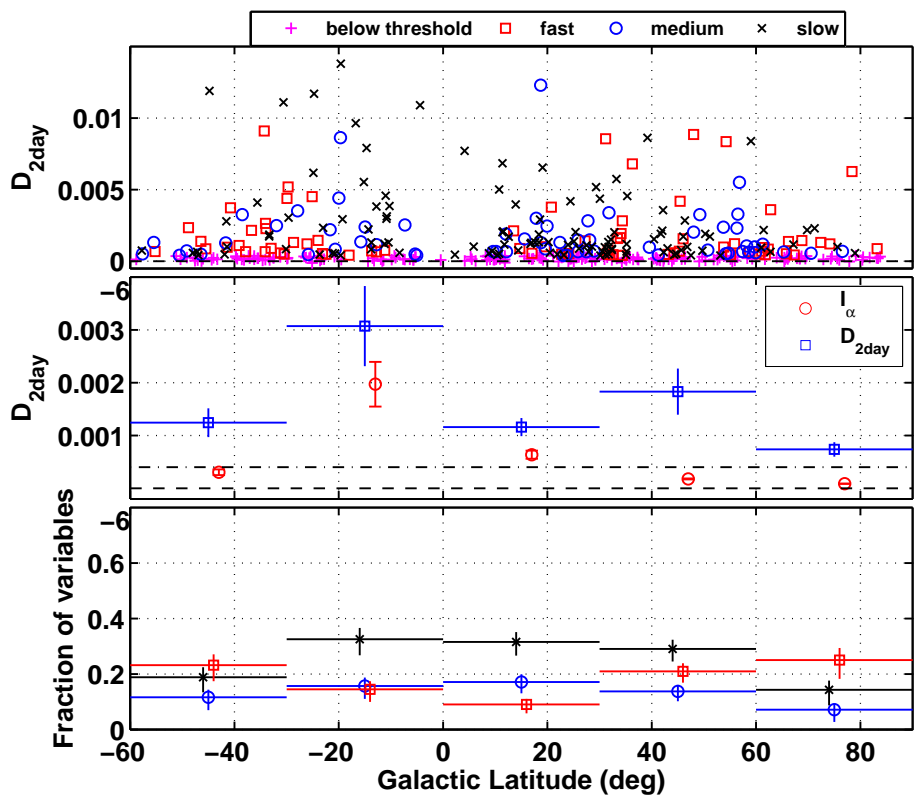} 
\caption{\it Upper panel: \rm Scatter plot of $D(2\mbox{d})$
(on log scale) against Galactic latitude. The intra-hour variable
source J1819+3845 is off scale at 0.25. The different symbols represent the
three classifications of timescale, as described in \S\ref{VarTimeScale}. 
\it Center panel: \rm Squares show mean value of $D(2\mbox{d})$ 
in 30 degree bins of latitude including all sources except J1819+3845; 
vertical bar gives the error in the mean. Note the N-S asymmetry in the
circles show mean H$\alpha$ emission in each bin. 
\it Lower panel: \rm  Fraction of sources above the threshold 
in each timescale class in each bin.
}
\label{fig:latplot}
\end{figure}

For the structure function analysis we simply plot $D(2\mbox{d})$
against the Galactic latitude of each source, in a fashion similar to
that of \citet{HR87}.  The upper panel of figure 
\ref{fig:latplot} is a scatter plot, differentiated by
the timescale group (fast, medium or slow). The middle panel averages 
$D(2\mbox{d})$ into 30 degree wide bins for all sources, which 
as already noted is independent of the threshold on $D(2\mbox{d})$. 
There is a low level of scintillation above 60 degrees, increasing 
in the mid range (30-60 deg), in both northern and southern hemispheres.
However in the low latitudes (0-30 deg) the ISS increases in the south 
of the plane but decreases north of the plane.  
While the figure is in reasonable agreement with the latitude 
dependence found by \cite{ric2006} in their analysis of the 
modulation index of 146 flat-spectrum sources observed at 2~GHz with the
Green Bank Interferometer, there are competing effects in our MASIV
survey since data were sampled for no more than 4 days.

As the latitude $b$ decreases both the distance $L$ to and the
path length through the scattering medium increase 
($\propto {\rm cosec}\;b)$.   The increased path 
length makes the scintillation stronger at lower latitudes
so that the modulation index should increase.  However, 
the increased distance $L$ increases the scale of the scintillation
pattern which slows the scintillation time
so that the structure function will saturate at times longer than 4 days,
causing a decrease in $D(2\mbox{d})$. The combination of these
two effects requires careful modeling.

The model for the structure function described in Appendix A
is a starting point for analyzing the effects of Galactic latitude.
Equation (\ref{eq:d2dmodel}) shows that as the distance $L$
increases $D(2\mbox{d})$ should decrease. However, the
equation assumes that the scintillation index of a point source
$m_{\rm pt}=1$, omitting any increase due to the longer scattered path length
at lower latitudes. In a more realistic model $m_{\rm pt}$ may be
less than one looking out of the Galactic plane ($|b| \sim 90$~deg)
and should increase with decreasing latitude, reaching unity
at $\sim\pm 45$~deg \citep{wal98}. At lower latitudes 
still it will increase only slowly due to effects of refractive ISS.  
The increase in $m_{\rm pt}$ between 90 and 45~deg partially compensates for
the reduction in $D(2\mbox{d})$ due to increasing distance $L$. At still 
lower latitudes $D(2\mbox{d})$ might be expected to decrease. 
In the observations one sees a difference of low latitude behavior
between the Northern and Southern hemispheres. 
This asymmetry can be understood by looking at the 
center panel of Figure \ref{fig:latplot} which shows
that the emission measure is commonly higher for Southern latitudes.

In a complete model the (unknown) compact fraction 
$f_c$ of the source flux density in the scintillating component
must also be considered.  Thus the expected variation of the ISS
level with Galactic latitude must be combined with the probability distributions 
for the flux fraction and for the diameters of the compact 
components,  which will further dilute
the variation of $D(2\mbox{d})$ with latitude.

In the Green Bank Interferometer observations cited above, the data were 
sampled daily (or on alternate days) over many years and so
provided estimates of the actual modulation index, which were not reduced
by the lengthening ISS time scale at low latitudes.  However,
even for these data the latitude dependence is not a strong effect.
Note that the asymmetry about the Galactic plane is very similar
to the asymmetry in the typical H$\alpha$ emission as shown by the
circles in the center plot.
The lower panel of figure \ref{fig:latplot} plots the 
fraction of sources in the three timescale
groups versus latitude and shows clear evidence that the fast scintillators
dominate at high latitudes and that slow scintillators dominate at 
low latitudes.  This agrees with the expected increase in ISS
timescale at low latitudes outlined above.

\subsection{Dependence of ISS on Source Spectral Index and Flux Density} 
\label{VarAlpha}

As stated earlier the ISS of extragalactic sources is expected to
be strongly suppressed, relative to that of pulsars, by the smoothing 
effect of their larger angular diameters.  Consequently, we expect
that the more compact sources should show higher levels of ISS.
Here we examine the influence of mean flux density ($\bar{S}$)
and spectral index $\alpha$ ($\bar{S} \propto \nu^{\alpha}$),
since we expect synchrotron emitting sources to be more compact for
larger $\alpha$ and lower $\bar{S}$, due to the effects of synchrotron
self-absorption and inverse Compton losses.
 
In \citet{hee84}'s initial survey of short-term variability of a 
large sample of both steep-spectrum and flat-spectrum sources he 
found that the flat-spectrum sources varied, ``flicker'', but 
the steep-spectrum sources did not. This can be understood as the 
steep-spectrum sources are dominated by optically thin synchrotron 
emission with low brightness temperatures, while the flat spectrum 
sources are dominated by synchrotron self-absorbed components with 
very high brightness temperatures \citep{SW68}, making them compact 
enough to show ISS.  

\begin{figure}[!ht]
\includegraphics[width=8cm,angle=270]{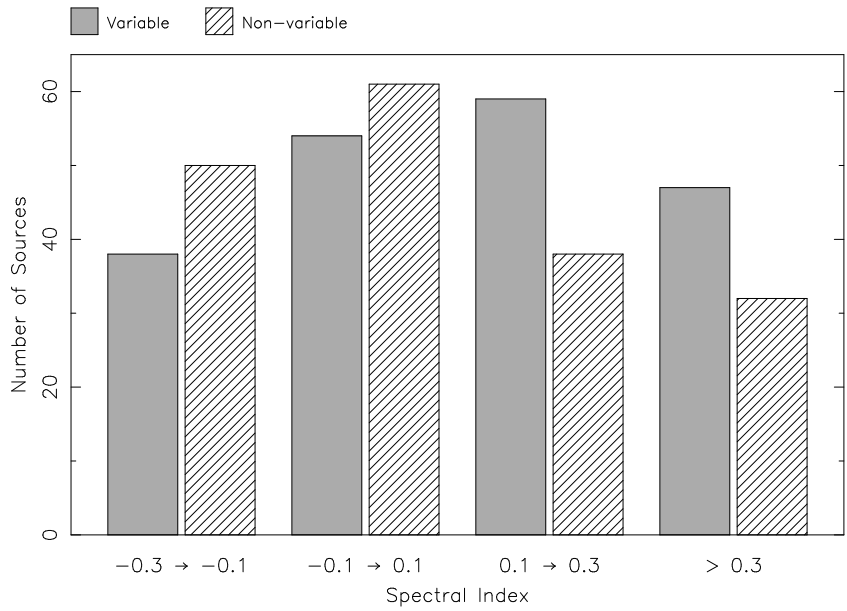}
\caption{Number of variable and non-variable sources as a function of spectral index.}
\label{fig:sidist}
\end{figure}

\begin{figure}[!ht]
\includegraphics[width=13cm]{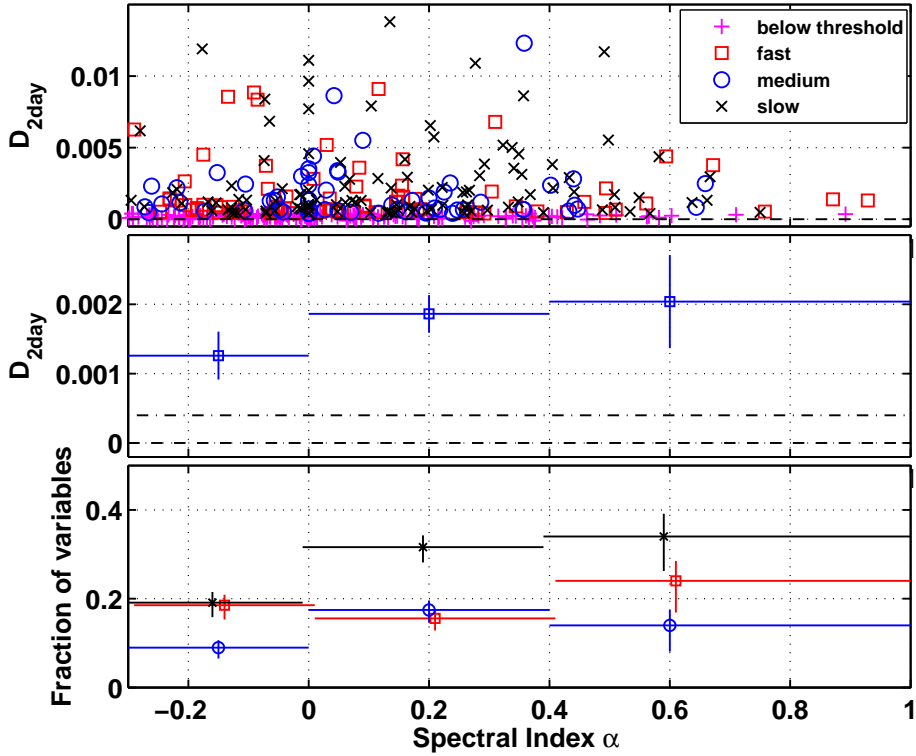}
\caption{$D(2\mbox{d})$  as a function of spectral index.
\it Upper panel: \rm Logarithmic scatter plot of $D(2\mbox{d})$. 
Different symbols represent the three classifications of ISS timescale.
\it Center panel: \rm Mean $D(2\mbox{d})$ binned by spectral index
including all sources except J1819+3845. \it Lower panel: \rm  Fraction of sources above the threshold in each timescale class in each bin.  
No significant trend is seen.
}
\label{fig:sid2d}
\end{figure}

The MASIV sources were selected to have flat spectral indices 
$\alpha > -0.3$  \citep{lov2003} and so are predominantly quasars with
compact cores.  However, it is possible that there are also 
some very compact galaxies in the sample.
Figure \ref{fig:sidist} shows the spectral index distributions
separately for the sources with the visual classification as 
variable or non-variable.  The spectral indices shown are those 
used to form the sample: 1.4 GHz NVSS \citep{con1998} to 8.5 GHz JVAS
\citep{pat1992,bro1998,wil1998} or
CLASS \citep{mye1995}  flux densities. This shows a slight increase in the fraction of sources that are variable with increasing $\alpha$, in agreement
with the expectation that the flatter (and inverted) spectrum sources are 
more compact. Though in Figure \ref{fig:sid2d} the mean $D(2\mbox{d})$
shows no significant trend with spectral index, the bottom panel
shows a slight increase in the fraction of variable sources
for $\alpha > 0$. This is mostly due to an 
increase in the fraction of slow variables which constitute the 
largest timescale group. It is worth pointing out that the surveys from which the flux densities were drawn to obtain spectral index were not coeval. It is likely then that any change in sample properties as a function of spectral index will be blurred as many of the sources vary intrinsically.

Turning to the influence of mean flux density, we first discuss the
visual variability classification and then plot 
$D(2\mbox{d})$ against flux density.  
The selection of sources for the MASIV survey divided them into
a high mean flux density group (strong) and a low mean flux density group 
(weak $S < 0.3$ Jy).
As reported in Paper 1 there was a greater fraction of variable 
sources in epoch 1 from the low flux density group than from the high flux density group.  Combining all four epochs the numbers of weak sources that varied in 0-4 epochs is (94, 46, 36, 39, 33) and for strong sources 
the numbers are (62, 45, 49, 22, 23).  Thus there are significantly 
more 3 \& 4 time variables among the weak sources than among the 
strong ones, though this trend is not supported in the 2 or 1 time variables.

\begin{figure}[!ht]
\includegraphics[width=13cm]{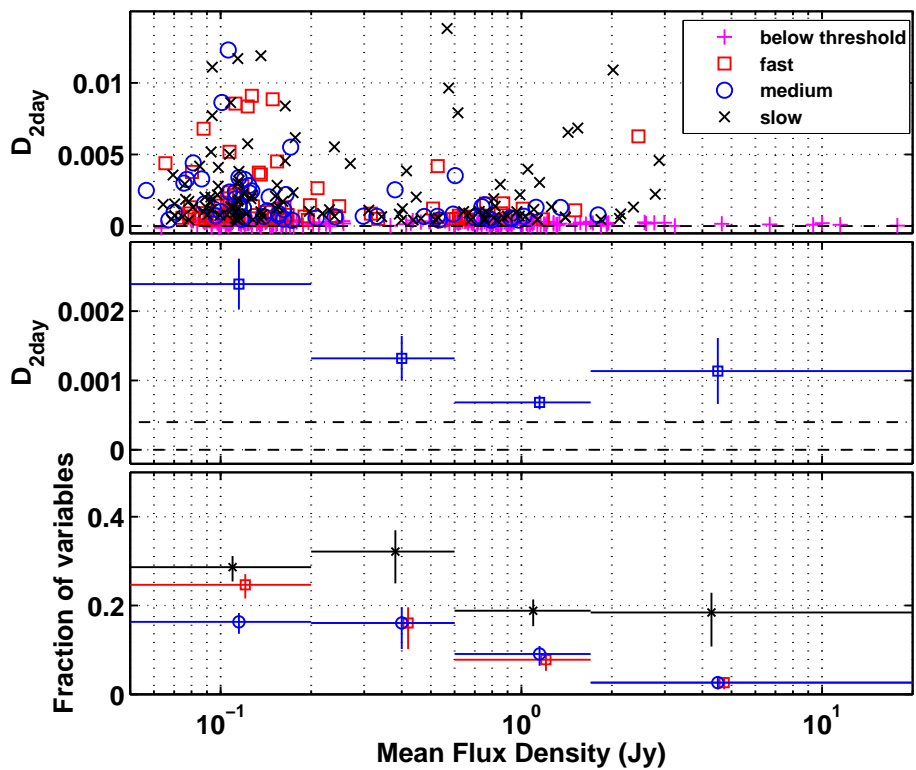} 
\caption{\it Upper panel: \rm Scatter plot of $D(2\mbox{d})$
against mean flux density. 
The higher values are more common
for lower flux density sources. The different symbols represent the
three classifications of ISS timescale, as described in the text.
\it Center panel: \rm Mean value of $D(2\mbox{d})$ 
in bins of mean flux density for all
sources (excluding extreme IHV quasar J1819+3845).
Note lower levels of ISS for the sources with higher flux density.  
The points centered at 2.5 Jy have only a few sources in each timescale
group giving larger errors in the mean for that bin.
\it Lower panel: \rm  Fraction of sources in each timescale class in each bin.
Note decreasing occurrence of fast variables among stronger sources. 
}
\label{fig:saveplot}
\end{figure}

Figure \ref{fig:saveplot} shows $D(2\mbox{d})$ versus 
mean flux density. The center panel shows a clear downward 
trend with increasing flux density in the lowest three bins,
while in the fourth bin it increases but with
fewer sources the mean has a large error.
The interpretation is an increasing angular diameter of the 
compact source components with increasing total mean flux density.  
Furthermore, the bottom panel shows a decrease in the fraction
of fast and medium scintillators with increasing mean flux density.
These are exactly the trends expected if their effective angular diameters 
are constrained by a maximum brightness temperature due to self-absorption
or inverse Compton losses ($\theta \propto (\bar{S}/T_{\rm B})^{0.5}$).

\subsection{Source Models}
\label{sec:sourcemodels}

The foregoing analysis establishes that about half of the
443 compact flat spectrum radio sources in the MASIV
survey show ISS at an rms level above 1\% over times of
2~d.  We now consider a simple model for the
compact source structure based on Appendix A.
Equation (\ref{eq:d2dmodel}) gives an approximate relation between 
$D(2\mbox{d})$ and parameters of the source
($f_c$ and $\theta_{\rm src}$) and of the interstellar medium
($L$ and $V$).  We proceed by assuming a basic model for the 
latter parameters $L=500$~pc and $V=50$~km~s$^{-1}$
and finding constraints on the source.  

\begin{figure}[!ht]
\includegraphics[width=15cm]{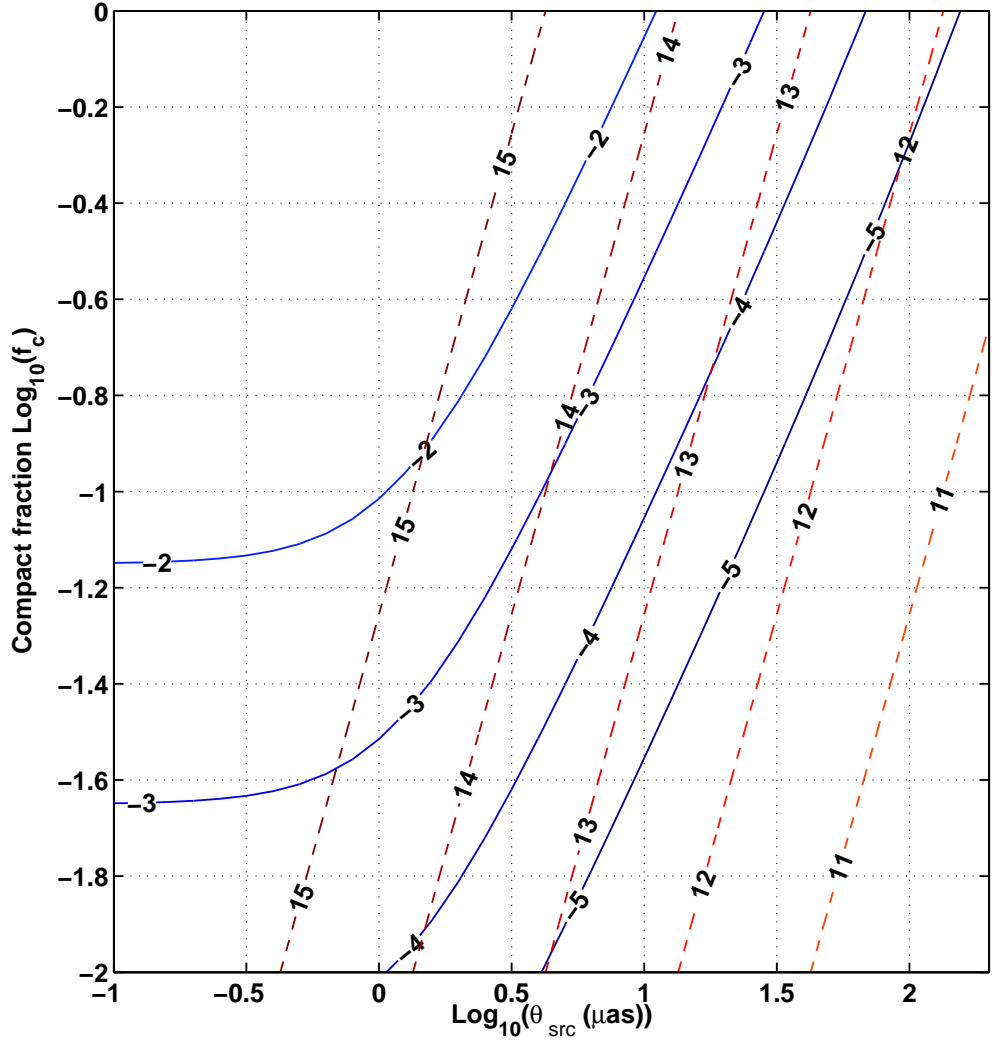} 
\caption{Solid lines show contours of $\log_{10}[D(2\mbox{d})]$ 
versus $f_c$ and $\theta_{\rm src}$ based on a Kolmogorov model for ISS
in a region at a distance of 500~pc.  Dashed lines are contours of
$\log_{10}[T_b/\bar{S}_{\rm Jy}]$ for total source flux density 
$\bar{S}_{\rm Jy}$.
}
\label{fig:d2dmod}
\end{figure}

Figure \ref{fig:d2dmod} shows (solid) contours of $D(2\mbox{d})$ versus 
the source parameters: compact
component diameter and compact flux fraction -- defined in the
observer's frame.
Also shown are (dashed) contours of $T_b/\bar{S}_{\rm Jy}$.
The majority of sources are in the range $0.0004 < D(2\mbox{d}) < 0.01$
which, for sources of 0.1 Jy mean flux density,
maps to maximum brightness temperatures $10^{12}$K to $10^{14}$K.
These figures require substantial Doppler factors in the AGN jets
comparable with those estimated from VLBI. 
However, we note that the plot of Figure \ref{fig:d2dmod} 
only provides a guide since it is based
on 500~pc as the distance to an interstellar scattering screen.

The distribution of scattering electrons along the line of
sight to each source is likely to be much more complex
and can extend from tens of pc to a few kpc.  Scattering at tens of
pc has been shown to be important for the rare rapid scintillators 
(IHV) and scattering from more than 1~kpc is responsible for the
slower timescale ISS associated from sight lines with large
emission measures.  Since the implied 
angular diameters scale roughly with the distance, the uncertainty in $L$
corresponds to an extremely large range in implied brightness temperatures.  
We note, however, the fast ISS sources in our sample are mostly
scintillating at levels of only 1-5\%, unlike the large amplitude
variations in the well studied IHV sources B0405--385, B1257--326 
and J1819$+$3845.  This suggests that the nearby scattering regions responsible
are relatively rare covering only a small fraction of the sky.
\cite{laz08} discuss the relative importance of sparsely distributed
``clumps of scattering material'' and a more uniformly distributed
interstellar scattering plasma, suggesting that the former could be more important for ISS and the latter for angular broadening in the ISM.
These authors find minimum diameters of $\sim1-2$~mas at 1~GHz,
which they suggest is caused by interstellar scattering, which
predicts $40-80 \mu$as when scaled to 5~GHz.  It will be important to 
use the full MASIV data set to re-examine these questions, but
since our emphasis here is on presenting the data, 
we postpone them to a later paper.

\subsection{Combined effects of Intrinsic Variations and ISS}

Here we examine the competing contributions that scintillation and
intrinsic variability would potentially make to the measured lightcurves.  
A source at an angular diameter distance $D_A$ undergoing intrinsic 
variations on a timescale $\tau$ has an implied maximum
intrinsic angular size:
\begin{eqnarray}
\theta = 17.3 \,{\cal D} \left( \frac{\tau}{100\,{\rm days}} \right) 
\left( \frac{D_A}{1\,{\rm Gpc}}\right)^{-1}\,\mu{\rm as}, 
\end{eqnarray}
where ${\cal D}$ is the Doppler factor. Further, a variation in flux 
density $\Delta S$ implies an observed brightness temperature of
\begin{eqnarray}
T_B = 6.4 \times 10^{12} \, \left( \frac{\Delta S}{100\,{\rm mJy}} \right) 
\left( \frac{\tau}{100\,{\rm days}} \right)^{-2} 
\left( \frac{D_A}{1\,{\rm Gpc}}\right)^{2} \, {\rm K}.
\label{eq:Tbobs}
\end{eqnarray}
When mapped into the emission rest frame the brightness temperature
is then reduced by a factor ${\cal D}^3 (1+z)^{-3}$,
under the hypothesis of intrinsic variation.

As an example consider a source that undergoes $100\,$mJy 
intrinsic fluctuations in $100\,$days, as observed between epochs for several
of our sources.  
At a typical distance $D_A\sim 1{\rm Gpc}$ the implied
maximum source size would be $\sim 17\,{\cal D} \mu$as.  
Suppose further that it does not show ISS within an epoch, which
implies it must be larger than $~80\,\mu$as in the observed frame
and so ${\cal D}>4.6$.  Hence mapping equation (\ref{eq:Tbobs}) 
into the emission frame gives $T_{\rm B,em} \simless 5\times 10^{10}$K.
We conclude that sources showing intra-epoch (intrinsic) variation and no ISS
have relatively low emission frame brightness
and, conversely, higher brightness sources that show intra-epoch 
variation have to show ISS.

\subsection{Dependence of ISS on Source Redshift} \label{VarZ}

We found redshifts for about half of the 443 sources in the survey 
from the published literature and we have subsequently measured
another 69 \citep{Purs08} for a total of 275 redshifts.  This
constitutes the largest sample of ISS measurements versus redshift.

\begin{figure}[bht]
\includegraphics[width=13cm]{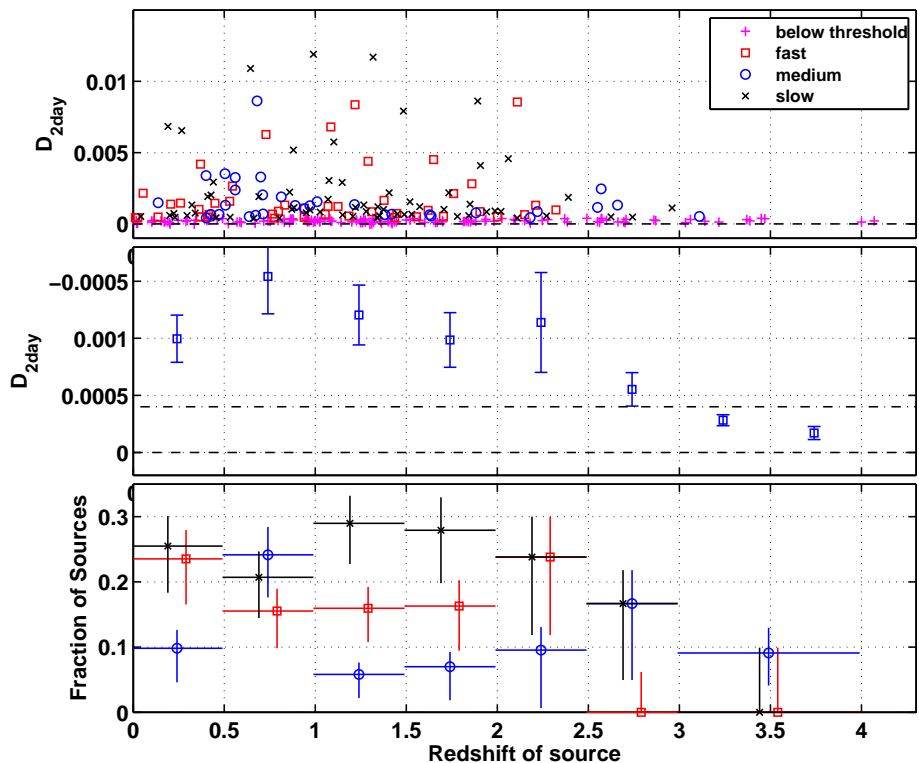} 
\caption{\it Upper panel: \rm Scatter plot of $D(2\mbox{d})$
against source redshift. The different symbols represent the
three classifications of ISS timescale, as described in the text.
\it Center panel: \rm Mean value of $D(2\mbox{d})$ 
in redshift bins for the 271 sources (out of 443) with measured
redshift (excluding extreme IHV quasar J1819+3845).
Note lower levels of ISS at high redshift.  
Values below the dash-dot line are upper bounds since they may be
raised slightly by low level confusion.
\it Lower panel: \rm  Fraction of sources in each timescale class in each bin.
The two bins above z=3 have been combined, as there are so few sources
at this redshift. 
}
\label{fig:zplot}
\end{figure}

Figure \ref{fig:zplot} plots $D(2\mbox{d})$ versus redshift and reveals a highly
significant decrease in the prevalence of ISS as redshift increases.  In particular
the middle panel shows that when binned in redshift 
the mean level of ISS drops steeply above redshift 2. 
As in other plots in this format the binned averages
are independent of any SF threshold. However, note that the exact value
of the lowest binned averages of $D(2\mbox{d})$ 
depend on the details of the noise subtraction.  
We note that, since the effect of confusion is to slightly increase
$D(2\mbox{d})$, our estimates become upper limits when 
below the threshold of about 0.0004.  

The bottom panel suggests that
the fraction of fast variables drops more quickly with redshift
than the fraction of slow and medium variables. 
However, the error bars show that this is only marginally significant,
and in view of the importance of this question we list 
the source counts in Table \ref{tab:redshiftcounts}. 
If true it would imply that the drop in mean ISS level is due to
an increase in angular diameter with redshift,
which also lengthens the ISS timescale. While the drop in ISS 
level seen in the middle panel could be due either to an increase in
diameter of the compact core of emission from these sources
or a decrease in the compact fraction of flux density in that core,
the latter interpretation would not explain a steeper decrease in
fast ISS with redshift than in slow ISS.  
The most likely conclusion from this analysis 
is that the extremely compact emitting regions 
responsible for the ISS in over half the quasars studied
appear broader in angular diameter with redshift above 2.
The interpretation of this result involves either an evolution
in the emitting regions with redshift or an angular broadening 
phenomenon due to propagation. We caution that a full consideration 
of selection effects must be made when interpreting this result. For 
example, redshifts are currently more complete in the strong (84\%) 
than the weak (43\%) sub-samples. Therefore, although this new 
cosmologically important result is the major finding
from the MASIV survey, we postpone a full discussion of 
the interpretation to a paper in preparation \citep{macq08}
pending a thorough investigation of selection effects \citep{Purs08}.  
Interested readers can consult preliminary discussions of the
interpretation by \citet{ricMRU}.
We also note that \cite{laz08} plotted angular diameter at 1~GHz
against redshift from a much smaller sample of scintillating and non-scintillating sources but could draw no firm conclusions.

\subsection{Intermittent Variability} \label{CompareVar}

We now ask whether the sources that were only classified as variable in
one to three of the four epochs are varying episodically or
are the result of statistical uncertainty and a fixed threshold for variability
on the raw modulation index.  As a result of the measurement uncertainties
there can be both false positives and false negatives, whose 
probabilities we can estimate.  Concentrating on the 91 
sources that were classified as variable in only one epoch
and correcting for false positives leaves us with 61 sources $\sim 13$\% of the
total.  These may be a category of short-lived, episodic
scintillators revealed by our regular sampling over a full year.

We note the strong case for intermittent ISS in the rapid variable 
PKS~0405--385 \citep{ked2001b}, and so consider a simple model for the intermittency in terms 
of the longevity of each IDV episode and the duration between episodes.
Intermittency may arise either due to fluctuations in the level of turbulence
responsible for ISS, or it may arise if the lifetimes of the 
bright, microarcsecond source components that undergo ISS are short.  
In the latter case one might expect sources to undergo IDV in conjunction 
with each outburst of the central engine.

Consider a simple model in which the IDV episodes have a finite duration 
$\Delta T$ and an interval, $T_f$, between outbursts.  Obviously in 
reality both quantities will have a distribution of possible values, 
but given the infrequency of our time sampling we restrict ourselves 
to this simple assumption here.  For any given source the
IDV episode commences at some random time $t_i \in [0,T_f)$, with the 
probability distribution of episodes distributed uniformly:
\begin{eqnarray}
p(t_i) = T_f^{-1}, \quad 0 \le t_i < T_f.
\end{eqnarray}
If we make a single observation the probability that the source will be exhibiting IDV is 
\begin{eqnarray}
p_1=\Delta T/T_f,
\end{eqnarray}
so for a survey that examines $N$ intermittent sources the expected 
number showing IDVs at any one time is $N \Delta T/T_f$. One can 
similarly calculate the probability of detecting IDV in a source 
during one or more of multiple observing epochs of a multi-epoch 
survey (see Appendix \ref{IntermittApp}).  In particular, the 
probability of detecting IDV in a source in one or more epochs 
of a four epoch survey, with epochs separated evenly in time by 
$t_{\rm obs}$, is 
\begin{eqnarray}
p_4 = \left\{ \begin{array}{ll}
\frac{\Delta T + 3 t_{\rm obs}}{T_f}, & t_{\rm obs} \leq \Delta T, \\
\frac{4 \Delta T}{T_f}, & t_{\rm obs} > \Delta T. \\
\end{array}
 \right. 
\end{eqnarray}
The number of IDV sources detected is a maximum when the interval 
between observing epochs exceeds the duration of IDV episodes because 
for shorter epoch intervals, after one merely discovers few new IDVs 
after the first epoch, one only keeps reobserving all the IDVs that
were present in the first epoch.  (Obviously, in the limit when 
$t_{\rm obs}$ is small, multiple observations discover the same 
number of sources as a single-epoch survey.)

Now, the mean detection rate of IDV sources in each epoch 
is 30\%, whereas the fraction of sources that exhibited IDV 
in one or more of our four epochs is 58\%.  These two numbers 
imply a typical burst duration $\Delta T = 1.2$\,yr and a 
burst period of $T_f = 3.8\,$yr.

We can also calculate the corresponding probability that IDV 
is observed in a source in all four epochs:
\begin{eqnarray}
p_{\rm all 4} = \left\{ \begin{array}{ll}
\frac{\Delta T-3 t_{\rm obs}}{T_f}, & 3 t_{\rm obs} < \Delta T, \\
0, & 3 t_{\rm obs} > \Delta T.
\end{array}
\right.
\end{eqnarray}
Based on the foregoing estimates of $\Delta T$ and $T_f$ one 
estimates that only 4\% of all sources should be common to all 
four epochs.  However, the actual detection fraction is 12\%.

It should be remembered that this model does not take 
into account several effects which are likely to be important, 
including: (i) annual cycle effects influence the number of 
sources one detects at any one epoch (ii) there is likely 
a distribution of episode durations and repetition rates, 
(iii) the repetition is likely irregular and (iv) not all 
IDVs are likely to be episodic.  We favor (ii), the importance of  
which is illustrated by the fact that many one-time 
IDVs were seen in epochs 2 \& 3 that were (obviously) 
not detected in epoch 4.  However, our model implies 
that sources that commenced IDV in these epochs should 
have been detected subsequently because the predicted 
burst duration exceeds the interval between observing 
runs (i.e. $\Delta T > 3 t_{\rm obs}$).

\section{Conclusions} 
\label{Conc}

We have reported results from the four epochs of the MASIV survey.  There were 
710 sources with flat spectra ($\alpha<-0.3$) near 5~GHz selected in 
weak and strong flux density groups surveyed for variability in four epochs
over a year.  These flat spectrum sources are predominantly quasars with
compact emission probably from a core and jet, many with effective diameters small
enough to show interstellar scintillation (ISS).  In each epoch the flux density was measured
using sub-arrays of the VLA every~2 hrs for about 12~hrs each day for 3-4~days.  
Sources were removed from the study if they showed evidence for
changing correlated flux density due to confusion or resolution of their more extended
structure, leaving 443 sources which were analyzed for variability in two
ways.  

The first was a binary classification based on the raw modulation index 
(visual method) in which 43\% of the 
sources were classified as variable in 2, 3 or 4 of the epochs. 
The second was a fit to the epoch-averaged structure function
parameterized by $D(2\mbox{d})$ and a timescale $\tau_{\rm char}$.
In view of the uncertainties in the latter
we classified sources as slow ($>3$~days), medium (0.5-3~days) or 
fast ($<0.5$~days) if $D(2\mbox{d})$ exceeded 
$4\times10^{-4}$.  By this criterion 37\% of the sources varied 
with more than 1.4\% modulation index over 2~days, which is similar
to the 43\% variables by the visual classification.

We found that $D(2\mbox{d})$ and timescale
varied both with coordinates in the Galaxy and also with source-based quantities.
This confirms that the variations are dominated
by ISS, which depends on both the strength of scattering and the distance
to the scattering region and also on the fraction of flux density 
in its most compact component and its effective angular diameter.  The following
is a summary of our findings:

\begin{itemize}

\item 
The amplitude of 2-d  variability increases with increasing emission 
measure estimated from H$\alpha$ intensity for each line of sight.  
Emission measure is the column density for the square of
the electron density which is expected to be strongly correlated with
inhomogeneity in the ionized medium that causes ISS. 
This result provides observational evidence that ISS is the dominant cause
of the variations.  We find fast variations dominate for 
low emission measure, as expected since such regions will be seen 
out of the plane and closer to the Earth, and
slow variations dominate for high emission measure which are typically
seen at greater distances toward the Galactic plane especially for
southern latitudes where the H$\alpha$ intensity is high.

\item 
The amplitude of 2-d ISS variability varies significantly with 
Galactic latitude but differs substantially between positive and negative
latitudes.  The expected behaviour is
complicated;  greater path lengths at low latitudes, where the 
scattering should be stronger, cause the scattering to be slower
which should reduce the rms over 3 d. 
However, the observed timescales show that there are more sources with
fast variations at high latitudes and more sources with
slow variations at low latitudes in both hemispheres, in
clear support of ISS as the dominant cause.

\item The ISS modulation index tends to decrease with increasing mean flux
density, as expected if the compact emission is limited
by synchrotron self absorption or inverse Compton losses to
have a maximum brightness temperature.  In that case the expected
angular diameter $\propto \sqrt{\bar{S}}$ which will quench the ISS
of the stronger sources. 

\item There is little change in the ISS amplitude with spectral index
for our sample with $\alpha > -0.3$. 

\item There is evidence that the ISS can be intermittent on times of 3-6 months
for some sources, but this is hard to quantify from the 
3~day observing sequences, when the time scale of
the variations is of the same order.  
 
\item  We model $D(2\mbox{d})$ as a function of compact source component
fractional flux density and angular diameter, from which we find 
compact diameter to lie in the range 0.005 -- 0.15 milli arcseconds
and brightness temperatures in the range $10^{12} - 10^{14}$K. 

\item
The most far-reaching result reported here is the discovery of
a decrease in both the fraction of sources that scintillate and in their 
scintillation amplitude beyond redshifts around 2.  We conclude
that there is an increase in the typical angular diameter
of the most compact radio-emitting regions of the quasars beyond reshift 2. 
The possible interpretations 
of this exciting result will be presented in a companion paper 
\citep{macq08}.

\item
A further surprise (at least to us) was the apparent absence of the
very rapid variables (IHV). J1819+3845 fell in our sample, but it was the
only source to show remarkable large amplitude variability. J0929+5013 showed rapid
variability in the January 2002 epoch \citep{lov2003} but,
although monitored closely, revealed only slower, many-hour
variability in the three later epochs. We had expected to find more of
these rapid variables especially given that two of the three known,
J1819+3845 and PKS1257-326 were discovered serendipitously.  This 
strongly suggests that the IHV sources lie behind discrete local
interstellar clouds which cover a small fraction of the sky.

\end{itemize}

\acknowledgements

The National Radio Astronomy
Observatory is a facility of the National Science Foundation operated
under cooperative agreement by Associated Universities, Inc. 
We are extremely grateful for the technical support provided by NRAO
staff at Socorro, in particular we would like to thank Ken Sowinski,
Miller Goss, Mark Claussen and Jim Ulvestad for helping to implement
the five subarrays at the VLA. 
This research has made use of the NASA/IPAC Extragalactic Database
(NED) which is operated by the Jet Propulsion Laboratory, California
Institute of Technology, under contract with the National Aeronautics
and Space Administration.  We also made use of data from the
Wisconsin H-Alpha Mapper, which is funded by the National Science Foundation.
BJR thanks the US-NSF for funding under grant AST 0507713.
He also thanks both the Cavendish Astrophysics group at Cambridge 
and the ATNF, Epping for hospitality.

\clearpage
\appendix 

\section{Structure function for ISS} 
\label{DsISS}

In all of the MASIV data the intrinsic source diameters
($\theta_{\rm src}$) are large enough to suppress the scintillations relative to
those of a point source (such as a pulsar), which at 6~cm would 
be expected to have a true modulation index 
$m_{\rm pt}$ near unity.  In this section we describe a model
for the structure function for such extended sources; 
this is easier to interpret than the apparent modulation index $m_i$.

The source diameter smoothes the ISS of a 
point source and so reduces the modulation index to $m_{\theta}$ 
and lengthens the timescale to $\tau_{\theta}$.  If the 
scattering is concentrated at a distance 
$L$ from the Earth and we are near the transition from 
weak to strong scintillation, a useful approximate relation is:
\begin{eqnarray}
m_{\theta} \sim m_{\rm pt} \frac{\theta_{\rm F}^{7/6}}{(\theta_{\rm F}^2 + 
\theta_{\rm src}^2)^{7/12}} \, .
\label{eq:mtheta}
\end{eqnarray}  
Here $\theta_{\rm F} = \sqrt{\lambda/(2 \pi L)}$ is the angular 
size subtended by the Fresnel scale
($r_{\rm F}$) and  $\theta_{\rm src}$ is the angular radius of the source.
The exponents $7/6$ and $7/12$ apply for a Kolmogorov 
spectrum of interstellar turbulence \citep{Coles87}. 
To the same accuracy the ISS timescale for a point source would be 
$\tau_{\rm F} = L\theta_{\rm F}/V$ and the source smoothing would increase it to:
\begin{eqnarray}
\tau_{\theta} = \tau_{\rm F} \frac{\sqrt{\theta_{\rm F}^2 + \theta_{\rm src}^2}}{\theta_{\rm F}} \, ,
\label{eq:tautheta}
\end{eqnarray}
where $V$ is the effective transverse velocity of the observer relative to the layer of scattering plasma.
Note that when the source diameter is sufficiently large to suppress the scintillations we have
\begin{eqnarray}
m_{\theta} \sim (\theta_{\rm F}/\theta_{\rm src})^{7/6} \; ,  
\tau_{\theta} \sim L \theta_{\rm src}/V \, ,
\label{eq:mtau}
\end{eqnarray}
where we have set $m_{\rm pt}=1$ \citep{R86}.

The theoretical form of the auto-correlation function for an extended source 
that substantially suppresses the scintillation index is given by the 
low wavenumber approximation of \citet{Coles87}.  This in turn gives 
the theoretical structure function, whose detailed shape depends on 
both the spectrum of the plasma density and on its distribution along 
the line of sight through the Galaxy.   Figure 14 of
\citet{ric2006} shows the form for sources with Gaussian brightness 
with peak temperature $10^{11}-10^{13}$K,  when the medium is modeled by the \cite{cl05}
electron distribution at a Galactic latitude of 45$^\circ$ 
(away from the Galactic center).  The form of the structure function 
at small time lags depends strongly on the density distribution in 
the local ISM.  A useful approximation to these results is given by:
\begin{eqnarray}
D(t) = 2 f_c^2 m_{\theta}^2 \frac{t^a}{t^a+\tau_{\theta}^a} \, ,
\label{eq:Dmodel}
\end{eqnarray}
where $1 \le a \le 2$ is a constant that depends on the density 
distribution in the local ISM.  Here $a \sim 2$ for a local bubble 
with low turbulence such that the effective scattering distance 
is beyond the bubble ($> 100$~pc) and alternatively $a \sim 1$ 
if the medium is uniformly turbulent out to a scale height 
(as in the ``disk'' of the CL05 electron density model).  
We have also introduced an extra variable $f_c$ that is 
the fraction of the source flux density in the bright (core) component.

Equations (\ref{eq:Dmodel}) and (\ref{eq:mtau}) thus provide a simple interpretation 
for our estimates of $D(t=2\mbox{d})$.  Inserting $m_{\theta}$ 
and $\tau_{\theta}$ from equation (\ref{eq:mtau}) we obtain:
\begin{eqnarray}
D(2\mbox{d}) =  2 f_c^2 [\frac{1}{1+ 2\pi L \theta_{\rm src}^2/\lambda}]^{7/6} 
\frac{1}{1+(L\theta_{\rm src}/V 2\mbox{d})^a } \, ,
\label{eq:d2dmodel}
\end{eqnarray}
An important feature of this result is that $D(2\mbox{d})$ decreases steeply with increasing $\theta_{\rm src}$ and so provides a sensitive measure of source diameter.  In estimating $D(2\mbox{d})$ for every source-epoch we set $a=1$, which allows us to estimate the scintillation timescale $\tau_{\theta}$ without having to also estimate the exponent $a$.  $D(2\mbox{d})$ can be converted to an effective 2-day modulation index by $m_{2\rm{d}} = \sqrt{0.5D(2\mbox{d})}$.  We note that $m_{2\rm{d}}$ can substantially exceed the apparent modulation index $m_i$ when the time scale is longer than about 2 days.

\section{A simple model for IDV intermittency} \label{IntermittApp}

Consider a model in which a source outbursts every duration $T_f$ in time and each IDV episode lasts $\Delta T$.  The initial outburst time is unknown, but its probability is evenly distributed in the interval $t_i \in [0,T_f)$: $p(t_i) = T_f^{-1}$.  Now consider a function $\bar f(t)= 1-[H(t) - H(t-\Delta T)]$, which assumes the value one whenever the source shows {\it no} IDV but the value zero when it is on.    

Thus the fraction of the time in which the source is off for initial burst durations between $t_i$ and $t_i+d\delta t_i$ is $\bar f(t_i) p(t_i) \delta t_i$.  Thus the probability that IDV is absent is 
\begin{eqnarray}
p_{\rm 1 off} &=& \int_0^{T_f} dt_i \bar f (t_i) p(t_i) \nonumber \\
&=& 1- \frac{\Delta T}{T_f}.
\end{eqnarray}
Thus the probability that the source is observed to exhibit IDV is $1-p_{\rm 1 off} = \Delta T/T_f$.

Now suppose we look for IDV at times $t=0, t_{\rm obs}, 2 t_{\rm obs}, 3 t_{\rm obs}$.  The fraction of the burst times between $t_i$ and $t_i +\delta t_i$ for which IDV is absent in all four observations is $\bar f(t_i) \bar f(t_i+t_{\rm obs}) \bar f(t_i+2 t_{\rm obs})\bar f(t_i+3 t_{\rm obs}) p(t_i) \delta t_i$.  If we assume that the repetition period exceeds the total duration of our observations (i.e. $T_f > 3 \,t_{\rm obs}$), the probability of observing no IDV over all four epochs is 
\begin{eqnarray}
p_{\rm 4 off} &=& \int_0^{T_f} dt_i \bar f (t_i) \bar f(t_i+t_{\rm obs}) \bar f(t_i+2 t_{\rm obs})\bar f(t_i+3 t_{\rm obs}) p(t_i) \nonumber \\
&=& \left\{ 
\begin{array}{ll}
1- \frac{\Delta T+3 t_{\rm obs}}{T_f}, & t_{\rm obs} \leq \Delta T, \\
1- \frac{4 \Delta T}{T_f}, & t_{\rm obs} > \Delta T.  
\end{array}
\right.
\end{eqnarray}
Thus the probability that IDV is observed in any one or more of these four epochs is 
\begin{eqnarray}
p_{\hbox{\footnotesize any of 4 on}} = 1-p_{\rm 4 off} = \left\{ 
\begin{array}{ll}
 \frac{\Delta T+3 t_{\rm obs}}{T_f}, & t_{\rm obs} \leq \Delta T, \\
  \frac{4 \Delta T}{T_f}, & t_{\rm obs} > \Delta T.  
\end{array}
\right.
\end{eqnarray}

We can similarly consider the probability of observing IDV in multiple observations by defining the function $f(t) = H(t) - H(t-\Delta T)$ which takes the value one whenever the IDV is on and zero otherwise.  The probability that IDV is observed in all four epochs is thus
\begin{eqnarray}
p_{\rm 4 on} &=& \int_0^{T_f} dt_if (t_i) f(t_i+t_{\rm obs}) f(t_i+2 t_{\rm obs}) f(t_i+3 t_{\rm obs}) p(t_i) \nonumber \\
&=& \left\{ 
\begin{array}{ll}
\frac{\Delta T-3 t_{\rm obs}}{T_f}, & 3t_{\rm obs} \leq \Delta T, \\
0, & T_f > 3t_{\rm obs} > \Delta T.  
\end{array}
\right.
\end{eqnarray}

\clearpage

\begin{table}[ht]
\caption{A Summary of the observation dates, durations and VLA array 
configurations for the first four MASIV epochs.}
\label{tab:obs}
\smallskip
\begin{center}
\begin{tabular}{cr@{,\,}lcc}
\tableline
\noalign{\smallskip}
Epoch Number & \multicolumn{2}{c}{Start Time (UT)} & Duration (h) & Array Configuration \\
\tableline
1 & 2002 January 19 & 19:14 & 72 & D $\rightarrow$ A \\
2 & 2002 May 9 & 08:02      & 72 & A $\rightarrow$ BnA \\
3 & 2002 September 13 & 17:50 & 96 & B $\rightarrow$ CnB \\
4 & 2003 January 10 & 16:21   & 72 & C $\rightarrow$ DnC \\
\tableline
\end{tabular}
\end{center}
\end{table}

\begin{table}[ht]
\caption{Observed numbers and percentages of sources classified 
as variable in the four epochs. Percentages are given of the 
443 sources observed, which excludes those used 
as calibrators in more than one epoch.  The predicted percentage 
of misclassifications assuming all sources were non-variable
(where $P=0.05$ is the probability of a single misclassification). 
}
\smallskip
\begin{center}
\begin{tabular}{ccccc}
\tableline
\noalign{\smallskip}
Number of & Observed number & Observed percentage  & Predicted fraction  & Predicted fraction  \\
variable epochs &  of sources & of sources & of misclassifications  & if none are variable   \\
\tableline
0 & 161 & 36\% & $(1-P)^4$     & 81.5\%   \\
1 & 90  & 21\% & $4 P (1-P)^3$   & 17.2\% \\
2 & 79  & 18\% & $6 P^2 (1-P)^2$ & 1.3\% \\
3 & 58  & 13\% & $4 P^3 (1-P)$   & 0.05\%  \\
4 & 55  & 12\% & $P^4$         & 0.0006\% \\
\tableline
\end{tabular} \label{tab:SourceCounts}
\end{center}
\end{table}

\begin{table}[ht]
\caption{Statistics of SF and visual variability classifications
for the 443 source sample.}
\begin{center}
\begin{tabular}{cccccc}
\tableline
No. epochs ``variable'' & No. of Sources & \% Sources & Mean $D(2\,\mbox{d})$ &
No. $>4\times10^{-4}$ & No. $<4\times10^{-4}$ \\
\tableline
0 & 161 & 36\% & 0.00024 & 27 & 134 \\
1 & 90 & 21\% & 0.00071 & 54 & 36 \\
2 & 79 & 18\% & 0.0011 & 64 & 15 \\
3 & 58 & 13\% & 0.0027 & 58 & 0 \\
4 & 55 & 12\% & 0.0071 & 55 & 0 \\
\tableline
\end{tabular} \label{tab:compvar}
\end{center}
\end{table}

\begin{deluxetable}{lllccccclc}
\tabletypesize{\scriptsize}
\rotate
\tablewidth{0cm}
\tablecaption{MASIV source variability characteristics. Column 4 gives
H$\alpha$ in Rayleighs for WHAM.
Columns 5-8 give mean flux density for epochs 1-4;  
Columns 9-12 give raw modulation index in \% for epochs 1-4;  
Column 13 $\sigma_{\rm err}$ as \%; Column 14 gives the visual classification per 
epoch for variable (y), non-variable (n) and non-variable source used as a secondary calibrator (c); 
Column 15 gives $D(2{\rm d})$ from fit to epoch average SF where error -1 flags a non-convergent fit;
Column 16 gives timescale classifications, 1:$\tau_{\rm char}<0.5$d,  2:0.5d$<\tau_{\rm char}<3$d,  3:$\tau_{\rm char}>3$d,
0: $D(2{\rm d})$ below threshold of $4\times10^{-4}$.} 
\tablehead{
\colhead{Name}           & 
\colhead{RA(J2000)}   &
\colhead{Dec(J2000)}  &
\colhead{H$\alpha$ R}  &
\colhead{$\bar{S_\nu}$(Jy)}    & 
\colhead{Mod index \% } &
\colhead{\%err} &
\colhead{Var}  &
\colhead{$D(2{\rm d})$}  &  
\colhead{tim} }
\startdata
J0005+3820 & 00:05:57.17 & 38:20:15.1&   2.9 &  0.61  0.57  0.43  0.47 &  1.1  1.1  1.2  1.5 &  0.57 & nnny &  +4.73e-04 $\pm$   1.4e-05 &   3 \\ 
J0007+5706 & 00:07:48.47 & 57:06:10.4&  10.8 &  0.18  0.20  0.17  0.17 &  2.1  0.9  1.1  1.0 &  0.88 & nnnn &  +2.91e-04 $\pm$   4.3e-05 &   0 \\ 
J0009+1513 & 00:09:43.47 & 15:13:37.9&   0.7 &  0.16  0.16  0.17  0.16 &  1.0  1.7  2.0  1.3 &  0.94 & nnnn &  +4.92e-04 $\pm$   5.5e-05 &   2 \\ 
J0010+1724 & 00:10:33.99 & 17:24:18.7&   0.8 &  0.85  0.83  0.81  0.81 &  0.8  1.0  0.9  0.5 &  0.53 & nnnn &  +1.07e-04 $\pm$   1.3e-05 &   0 \\ 
J0010+4412 & 00:10:30.04 & 44:12:42.4&   2.3 &  0.20  0.20  0.19  0.18 &  1.5  1.2  1.5  1.7 &  0.83 & nynn &  +4.28e-04 $\pm$   3.9e-05 &   1 \\ 
J0011+7045 & 00:11:31.90 & 70:45:31.6&   6.7 &  0.62  0.71  0.63  0.59 &  1.9  0.6  0.6  1.4 &  0.54 & ynny &  +3.46e-04 $\pm$   3.6e-05 &   0 \\ 
J0017+5312 & 00:17:51.75 & 53:12:19.1&  13.3 &  0.61  0.60  0.61  0.65 &  1.5  0.9  1.2  1.2 &  0.54 & nnyy &  +3.88e-04 $\pm$  -1.0e+00 &   0 \\ 
J0017+8135 & 00:17:08.47 & 81:35:08.1&   2.2 &  0.91  0.92  0.92  0.92 &  1.4  1.2  0.5  0.7 &  0.52 & nnnn &  +2.13e-04 $\pm$   1.6e-05 &   0 \\
\enddata

\label{tab:BigTable}
\end{deluxetable}

\begin{table}[ht]
\caption{Counts versus redshift for sources with $D(2{\rm d})$ above the
threshold $4\times 10^{-4}$ .}
\begin{center}
\begin{tabular}{lccccccc}
\tableline
Redshift & 0-0.5 & 0.5-1 & 1-1.5 & 1.5-2.1 & 2.1-2.6 & 2.6-3.1 & 3.1-4.1 \\
\tableline
No. of sources  & 51 & 58 & 69 & 43 & 21 & 18 & 11\\
Fast & 12 & 9 & 11 & 7 & 5 & 0 & 0 \\
Medium & 5 & 14 & 4 & 3 & 2 & 3 & 1 \\
Slow & 13 & 12 & 20 & 12 & 5 & 3 & 0 \\
\% above threshold& 59 & 60 & 51 & 57 & 57 & 33 & 9 \\
\tableline
\end{tabular} \label{tab:redshiftcounts}
\end{center}
\end{table}

\end{document}